\documentclass[useAMS,usenatbib]{mn2e}

\usepackage{amsmath}
\usepackage{amssymb,graphicx}
\usepackage{verbatim}
\usepackage{journal_names}
\usepackage{pdflscape}

\title[A multi-wavelength survey of NGC\,6752]
{A multi-wavelength survey of NGC\,6752: X-ray counterparts, two new dwarf novae, and a core-collapsed radial profile}
\author[G.~S.~Thomson et al.]
{G.~S.~Thomson$^1$, C.~Knigge$^1$, A.~Dieball$^1$, T.~J.~Maccarone$^1$, A.~Dolphin$^2$, D.~Zurek$^3$, \and K.~S.~Long$^4$, M.~Shara$^3$, and A.~Sarajedini$^5$\\
$^1$Faculty of Physical and Applied Sciences, University of Southampton, SO17~1BJ, UK\\
$^2$Raytheon Company, 1151 E. Hermans Road, Tucson, AZ 85706, USA\\
$^3$Department of Astrophysics, American Museum of Natural History, New York, NY 10024, USA\\
$^4$Space Telescope Science Institute, Baltimore, MD 21218, USA\\
$^5$Department of Astronomy, University of Florida, Gainesville, FL 32611, USA\\}

\begin{document}
\pagerange{\pageref{firstpage}--\pageref{lastpage}} \pubyear{2012}
\maketitle
\label{firstpage}
\begin{abstract}
We present the results of a multi-wavelength (FUV to I-band) survey of the stellar populations of the globular cluster NGC\,6752, using STIS, ACS and WFC3 on board the \textit{Hubble Space Telescope}. We have confirmed that two previously identified CV candidates are, in fact, dwarf novae which underwent outbursts during our observations. We have also identified previously unknown optical counterparts to two X-ray sources. We estimate the position of the centre of the cluster, and show that the stellar density profile is not well described by a single King model, indicating that this cluster is in a core-collapsed or post-core collapse phase. The colour-magnitude diagram shows a well-populated horizontal branch, numerous blue stragglers and white dwarfs (WDs), as well as 87 sources in the gap region where we expect to find WD - main sequence binaries, including cataclysmic variables (CVs). The X-ray sources and WD binary systems are the most centrally concentrated populations, with dynamically estimated characteristic masses $>1.1\,M_{\odot}$ and $>0.8\,M_{\odot}$, respectively.
\end{abstract}


\section{Introduction}

\begin{table*}
\caption{Summary of observations used in this survey. The individual images from the FUV and NUV wave-bands are listed, as these were used to search for variability. In all other cases, total exposure times are listed.}
\label{obstable}
\begin{center}
\begin{tabular}{lcccccl}
\hline
\hline
Instrument/ & FoV                          & Plate Scale       & Waveband & Filter & Exposures & Date \\
Detector    &                              & [$\arcsec$/pixel] &          &        &           &      \\
\hline
STIS        & $25\farcs1\times25\farcs3$   & $0.25$ & FUV      & F25QTZ & $10\times650$\,s & 2001 March 30  \\
            &                              &        &          &        & $3\times900$\,s  & 2001 March 30  \\
WFC3/UVIS   & $162\arcsec\times162\arcsec$ & $0.04$ & NUV      & F225W  & $6\times120$\,s  & 2010 July 31   \\
            &                              &        &          &        & $6\times120$\,s  & 2010 August 7  \\
            &                              &        &          &        & $6\times120$\,s  & 2010 August 21 \\
WFC3/UVIS   & $162\arcsec\times162\arcsec$ & $0.04$ & U        & F390W  & $1590$\,s        & 2010 May 1 - 5 \\
WFC3/UVIS   & $162\arcsec\times162\arcsec$ & $0.04$ & B        & F410M  & $1800$\,s        & 2010 May 1 - 5 \\
ACS/WFC     & $202\arcsec\times202\arcsec$ & $0.05$ & V        & F606W  & $142$\,s         & 2006 June 24   \\
ACS/WFC     & $202\arcsec\times202\arcsec$ & $0.05$ & I        & F814W  & $162$\,s         & 2006 June 24   \\
\hline
\end{tabular}
\end{center}
\end{table*}

Globular clusters (GCs) have high stellar densities, especially towards the cluster core. This makes them ideal locations for the dynamical formation of exotic stellar populations such as X-ray binaries, milli-second pulsars (MSPs), blue stragglers (BSs), cataclysmic variables (CVs) and other close binary systems (e.g. Knigge et al. 2008; Dieball et al. 2005). Knowledge of the properties of the binary fraction of a cluster can help us to understand GC evolution, since interactions between binary systems and passing single stars can actually drive the cluster towards evaporation. Thus, GCs are not only excellent places to study close binaries, but studying the close binary populations also helps us to understand more about the GC itself.

Ultraviolet (UV) images are useful in searches for exotic stellar populations, because main sequence (MS) and red giant (RG) stars which dominate in optical images are fainter in the UV, so the crowding that affects optical images is no longer a problem. The most exotic populations, on the other hand, emit much of their radiation in the UV, making UV images an ideal tool to search for interacting binary systems.

NGC\,6752 is a nearby, dense GC, which lies at a distance of 4\,kpc, and has a reddening of $E_{B-V}=0.04$\,mag, and a metallicity of $[Fe/H]=-1.54$\,dex (Harris 1996, 2010 edition). In a study using $Chandra$ observations, Pooley et al. (2002) identified 19 X-ray sources in NGC\,6752 and found 12 optical counterparts, including 10 CV candidates.

D'Amico et al. (2002) then identified 5 MSPs, four of which have no known optical counterparts. The other (PSR\,A), has a white dwarf (WD) counterpart (Bassa et al. 2003), but this MSP is located $\approx74$ core radii from the centre, and it is not clear whether or not this source is a cluster member (Bassa et al. 2006, Cocozza et al. 2006). If PSR\,A is a cluster member, one of the likely explanations for its offset position is that it was propelled there by an interaction with a black hole-black hole binary system (Colpi, Possenti \& Gualandris 2002). This idea is supported by the unusually high mass-to-light ratio in NGC\,6752, suggesting an excess of low-luminosity stars in the core (D'Amico et al. 2002). Further evidence for the existence of black holes (BHs) in GCs might be found in the stellar radial distributions. A central BH is likely to produce a central cusp in the surface brightness profile, which can be distinguished from core collapse by its slope (Baumgardt et al. 2005).

Despite the fact that NGC\,6752 is a relatively close GC, there is no consensus on its dynamical status. There has been much discussion in previous studies regarding whether or not the GC should be classified as core-collapsed. Ferraro et al. (2003) found that the radial profile can only be modelled using a combination of two King (1966) profiles, which they interpret as indication that it has undergone core collapse, but other studies (e.g. Lugger, Cohn \& Grindlay 1995), argued that it is not inconsistent with a single King model. Noyola \& Gebhardt (2006) produced surface brightness profiles for 38 GCs and found that NGC\,6752 was the only GC previously reported as core-collapsed that did not show a central cusp. Like Lugger et al. (1995), they found that the central part of the surface brightness profile was flat.

Here, we present the results of a study performed using near ultraviolet (NUV), U- and V-band data taken with the Wide Field Camera 3 (WFC3) on board the \textit{Hubble Space Telescope} (\textit{HST}), as well as far ultraviolet (FUV) observations using the Space Telescope Imaging Spectrograph (STIS), and the V- and I-band catalogue from The ACS Survey of Galactic Globular Clusters (Sarajedini et al. 2007). Our goals include searching for counterparts to X-ray sources and identifying new interacting binaries. We will also investigate the puzzling dynamical status of the cluster by constructing a radial density profile and determining if it can be fit with a single King profile.

The paper is structured as follows. In Section \ref{obscat} we describe the observations and data reduction. In Section \ref{cmd} an analysis of the colour-magnitude diagram (CMD) is presented. In Sections \ref{xray}, \ref{msp}, and \ref{variables}, we discuss the counterparts to X-ray and radio sources, and the search for variable sources in our catalogue. In Section \ref{dynamical} we present a new estimate for the position of the cluster centre and investigate the dynamical status of the cluster as a whole. In Section \ref{stellarpops} we examine the distributions of various stellar populations and estimate their characteristic mass. Our conclusions are summarised in Section \ref{conc}.

\section{Observations and Catalogue}
\label{obscat}

\subsection{Observations}
\label{obs}

The data used in this survey come from three sources, summarised in Table \ref{obstable}. The majority of this paper deals with observations carried out using the UVIS detector on WFC3 on board \textit{HST}, with the F225W (NUV) and F390W (U-band) filters. The NUV data consist of 18 individual images of 120\,s exposure time, which are used in the search for time variability. The U-band dataset comprised 6 images with exposure times between 2\,s and 880\,s each, giving a total exposure time of 1590\,s. We also include 3 exposures with the F410M filter with exposure times of 40\,s and 880\,s. These data are included in the discussion of optical counterparts to X-ray sources, for completeness. WFC3 has a field of view of $162\arcsec\times162\arcsec$ and a plate scale of $0\farcs04$/pixel.

Secondly, we used data from the Advanced Camera for Surveys (ACS) in the Wide Field Channel (WFC), which have total exposure times of 142\,s with the F606W (V-band) filter and 162\,s with the F814W (I-band) filter (PI. Sarajedini). The ACS has a larger field of view than WFC3, at $202\arcsec\times202\arcsec$, but a slightly coarser plate scale of $0\farcs05$/pixel.

Finally, we used FUV observations consisting of 13 individual exposures of 650\,s or 900\,s taken on 2001 March 30 with STIS on board \textit{HST} using the F25QTZ filter. This has a wavelength range of 1475~-~1900\,\AA. The STIS observations have a field of view of $25\farcs1 \times 25\farcs3$ and a plate scale of $0\farcs025$/pixel. This dataset was used to facilitate time variability studies of sources in the core of the cluster.

\subsection{Photometry}
\label{phot}

\begin{figure*}
\includegraphics[width=0.6\textwidth]{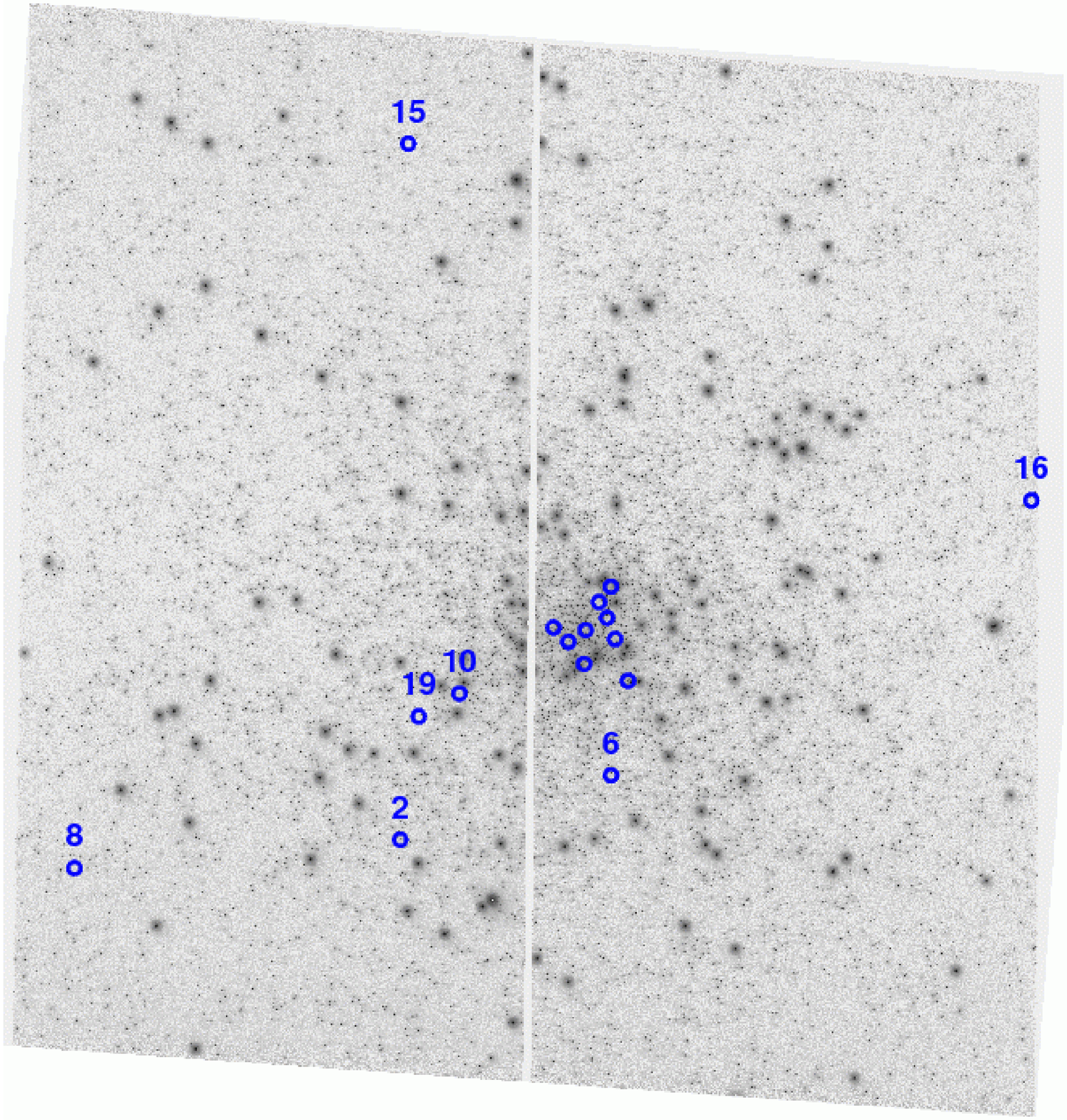}
\includegraphics[width=0.35\textwidth]{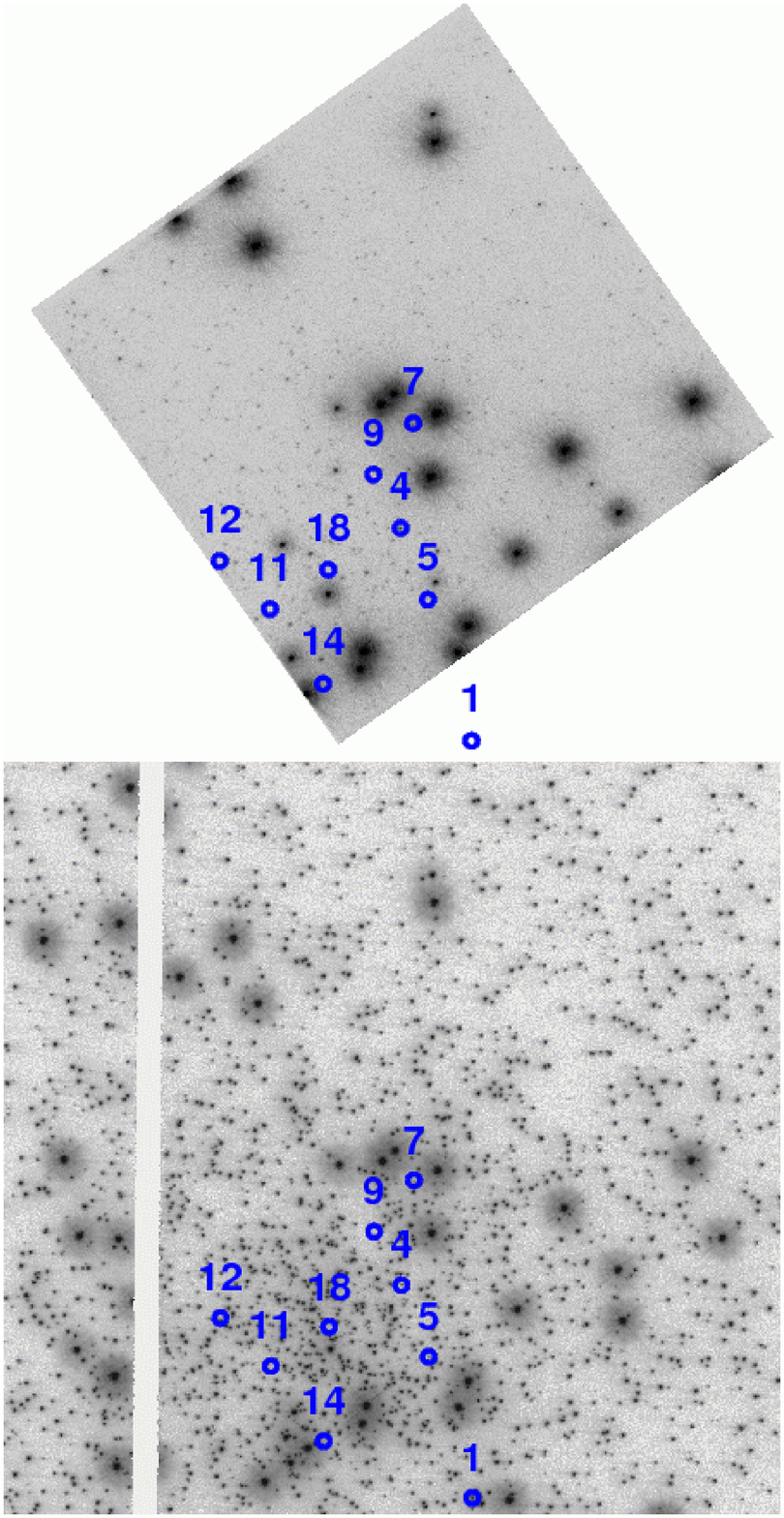}
\caption[fig_x_all]{Left panel: Combined and geometrically corrected `master' image of 15 NUV images. North is up and East is to the left. The field of view is $162\arcsec\times162\arcsec$ and the pixel scale is $0\farcs04$/pixel. X-ray source positions are marked in blue circles and labelled with their CX number (except those in the core). Top right panel: Combined `master' image of 13 FUV images, with X-ray positions marked. Again, North is up and East is to the left. The STIS field of view is $25\farcs1 \times 25\farcs3$ and the pixel scale is $0\farcs025$/pixel. Lower right panel: Zoomed in version of the central region of the master NUV image, with the central X-ray sources marked and labelled with their CX number. The field shown is $\approx40\arcsec$ wide and $38\arcsec$ high and the position and scale corresponds to that of the FUV image shown above.}
\label{fig_x_all}
\end{figure*}

Astrometry and photometry were performed on the WFC3 data (NUV, U- and B-band) using the WFC3 module of DOLPHOT (Dolphin 2000). First, master images to be used as reference images for the World Coordinate System (WCS) were made for each of the WFC3 datasets (one per filter), by combining the individual frames using {\tt multidrizzle} running under {\tt PyRAF}\footnote{Note that we are limited to combining a maximum of 15 frames by the capabilities of the \textit{wfc3mask} routine of DOLPHOT, but this only affects the NUV dataset as the U- and B-band data are made up of fewer individual exposures; this limit will not affect the outcome of the photometry.}. The DOLPHOT procedure \textit{wfc3mask} was used to mask the pixels identified as saturated or bad in the \textit{HST} pipeline. The DOLPHOT tasks \textit{splitgroups} and \textit{calcsky} were used to divide the exposures into their component chips (so they can be aligned to the master image), and to create a sky image for each frame. To correct for changes in alignment between individual frames, a few reference stars were found in each individual image and the reference image, and \textit{wfc3fitdistort} was used to determine how each individual frame differed from the reference image. DOLPHOT was then used to perform photometry on each individual image, giving a catalogue in the WCS of the reference frame. We cleaned the resulting catalogues by removing sources which were deemed too sharp or too extended, as these are likely to be cosmic rays or background galaxies, or which were badly affected by the presence of nearby neighbours so that PSF-fitting could not be completed adequately. The resulting catalogues contained 14511 NUV, 27099 U-band, and 32780 B-band sources.

The V- and I-band ACS data were taken from the ACS Survey of Galactic Globular Clusters (Sarajedini et al. 2007, Anderson et al. 2008; henceforth `the ACS Survey Catalogue'). The catalogue contains 52818 stars that were found in both the V- and I-bands.

The individual FUV exposures were aligned by hand using the {\tt IRAF}\footnote{{\tt IRAF} (Image Reduction and Analysis Facility) is distributed by the National Astronomy and Optical Observatories, which are operated by AURA, Inc., under cooperative agreement with the National Science Foundation.} routine {\tt imalign}, as image distortion coefficients are not available for these images, and then combined using the {\tt IRAF} routine {\tt imcombine}. We performed astrometry and photometry on the FUV images broadly following the method described in Dieball et al. (2010) and Dieball et al. (2007). We used the {\tt IRAF} routine {\tt daofind} (Stetson 1991) to create an initial source list, and then inspected the image by eye to add sources missed by {\tt daofind} and remove false detections. This left 503 sources in the FUV catalogue. Photometry was performed using the {\tt IRAF} routine {\tt daophot} (Stetson 1991). Due to the high stellar density in the core, we chose a small aperture of 3 pixels and a sky annulus of 5-7 pixels. We also used a few isolated, bright stars to determine corrections for the fact that some source flux will be included in the sky annulus, and for the finite aperture size.

\begin{table*}
\caption{Catalogue of all sources in our WFC3 field of view. The first column is the source ID number. Columns 2-5 give the source position in RA and decl. and image pixel coordinates (using our F225 `master' image). Column 6 is the FUV magnitude measured using {\tt daophot}. Columns 7 and 9 give estimates of the FUV and NUV variability amplitude, defined to be the standard deviation of the source relative to its mean magnitude. Columns 8, 10 and 11 give the NUV, U- and B-band magnitudes derived using DOLPHOT. Columns 12 and 13 give the corresponding V- and I-band magnitudes from the ACS Survey Catalogue. The final column shows the source type found using the NUV~-~U or V~-~I CMDs and any further comments. Only 20 entries are listed here to demonstrate the catalogue's form and content. A machine-readable version of the full table is available in the online version.}
\label{cattable}
\begin{center}
\scriptsize
\begin{tabular}{cccccccccccccl}
\hline
\hline
1    & 2            & 3            & 4        & 5        & 6       & 7             & 8      & 9           & 10     & 11     & 12     & 13     & 14       \\
\hline
ID   & RA           & Decl.        & x$_{NUV}$& y$_{NUV}$& FUV     & $\sigma$FUV   & NUV    & $\sigma$NUV & U      & B      & V      & I      & Comments \\
     & (hh:mm:ss)   & (deg:mm:ss)  & (pixels) & (pixels) & (mag)   & (mag)         & (mag)  & (mag)       & (mag)  & (mag)  & (mag)  & (mag)  &          \\
\hline
8001 & 19:10:52.440 & -59:59:05.81 & 2439.584 & 2354.241 &  ...    & ...           & 18.929 & 0.060       & 16.642 & 16.503 & 16.451 & 16.804 & RGB      \\
8002 & 19:10:52.296 & -59:59:05.88 & 2439.486 & 2381.566 &  ...    & ...           & 20.366 & 0.101       & 18.751 & 18.594 & 18.742 & 19.153 & MS       \\
8003 & 19:10:57.678 & -59:59:03.12 & 2439.706 & 1360.024 &  ...    & ...           & 22.662 & 0.310       & 19.868 & 19.706 & 19.567 & 19.916 & MS       \\
8004 & 19:10:52.072 & -59:59:06.01 & 2439.866 & 2424.109 &  24.014 & 1.002         & 19.639 & 0.051       & 18.256 & 18.267 & 18.376 & 18.855 & MS       \\
8005 & 19:10:52.136 & -59:59:06.01 & 2440.693 & 2412.017 &  ...    & ...           & 19.458 & 0.093       & 18.006 & 17.886 & 18.050 & 18.515 & MS       \\
8006 & 19:10:49.722 & -59:59:07.26 & 2441.015 & 2870.224 &  ...    & ...           & 20.386 & 0.061       & 18.746 & 18.655 & 18.770 & 19.229 & MS       \\
8007 & 19:10:51.289 & -59:59:06.46 & 2441.085 & 2572.808 &  ...    & ...           & 20.191 & 0.077       & 18.632 & 18.526 & 18.694 & 19.105 & MS       \\
8008 & 19:10:52.862 & -59:59:05.66 & 2441.262 & 2274.257 &  ...    & ...           & 18.476 & 0.043       & 17.156 & 17.062 & 17.308 & 17.802 & MS       \\
8009 & 19:10:47.418 & -59:59:08.47 & 2441.816 & 3307.576 &  ...    & ...           & 18.532 & 0.037       & 17.385 & 17.266 & 17.513 & 18.039 & MS       \\
8010 & 19:10:51.232 & -59:59:06.53 & 2442.112 & 2583.698 &  ...    & ...           & 19.567 & 0.044       & 18.265 & 18.126 & 18.323 & 18.812 & MS       \\
8011 & 19:10:49.680 & -59:59:07.34 & 2442.489 & 2878.296 &  ...    & ...           & 19.871 & 0.061       & 18.441 & 18.323 & 18.490 & 18.974 & MS       \\
8012 & 19:10:53.602 & -59:59:05.33 & 2442.524 & 2133.886 &  ...    & ...           & 17.912 & 0.026       & 14.248 & 14.036 & 13.648 & 13.881 & RGB      \\
8013 & 19:10:52.486 & -59:59:05.91 & 2442.698 & 2345.726 &  21.428 & 0.174         & 21.685 & 0.187       & ...    & ...    & ...    & ...    & MS       \\
8014 & 19:11:00.147 & -59:59:01.97 & 2442.849 & 891.5843 &  ...    & ...           & 18.327 & 0.031       & 16.869 & 16.726 & 16.889 & 17.347 & SGB      \\
8015 & 19:10:51.167 & -59:59:06.59 & 2442.784 & 2596.081 &  ...    & ...           & 18.238 & 0.042       & 16.872 & ...    & 16.961 & 17.424 & SGB      \\
8016 & 19:10:51.919 & -59:59:06.21 & 2442.926 & 2453.358 &  ...    & ...           & 20.415 & 0.112       & 18.811 & 18.685 & 18.750 & 19.214 & MS       \\
8017 & 19:10:50.705 & -59:59:06.83 & 2442.865 & 2683.775 &  ...    & ...           & 21.784 & 0.277       & 19.379 & 19.23  & 19.207 & 19.608 & MS       \\
8018 & 19:10:49.136 & -59:59:07.64 & 2443.033 & 2981.585 &  ...    & ...           & 23.535 & 0.966       & 20.189 & 19.962 & 19.801 & 20.146 & MS       \\
8019 & 19:10:51.987 & -59:59:06.19 & 2443.301 & 2440.477 &  22.124 & 0.312         & 18.106 & 0.055       & 16.841 & 16.8   & 17.275 & 17.818 & SGB      \\
8020 & 19:10:55.753 & -59:59:04.26 & 2443.436 & 1725.670 &  ...    & ...           & 19.072 & 0.044       & 17.782 & 17.663 & 17.936 & 18.435 & MS       \\
\hline
\end{tabular}
\normalsize
\end{center}
\end{table*}

\subsection{Matching the Catalogues}
\label{matching}

\begin{table}
\caption{Number of matches between catalogues. The first two columns indicate the datasets being matched. The third column gives the number of matches found. The last two columns give the expected percentage of false matches and the number of expected false matches that this equates to. See text for details.}
\label{matchtable}
\begin{center}
\begin{tabular}{llrcr}
\hline
\hline
\multicolumn{2}{c}{Catalogues} & N$_{match}$ & \%$_{false}$ & N$_{false}$ \\
\hline
STIS/FUV & WFC3/NUV            & 492         & 3.84         & 19\\
WFC3/NUV & WFC3/U              & 12020       & 0.05         & 6\\
WFC3/NUV & WFC3/B              & 11883       & 0.06         & 7\\
WFC3/NUV & ACS/V\&I            & 13494       & 0.03         & 4\\
WFC3/U   & WFC3/B              & 22910       & 0.08         & 18\\
WFC3/U   & ACS/V\&I            & 24258       & 0.03         & 7\\
WFC3/B   & ACS/V\&I            & 23045       & 0.16         & 37\\
\hline
\end{tabular}
\end{center}
\end{table}

\begin{figure*}
\includegraphics[width=0.95\textwidth]{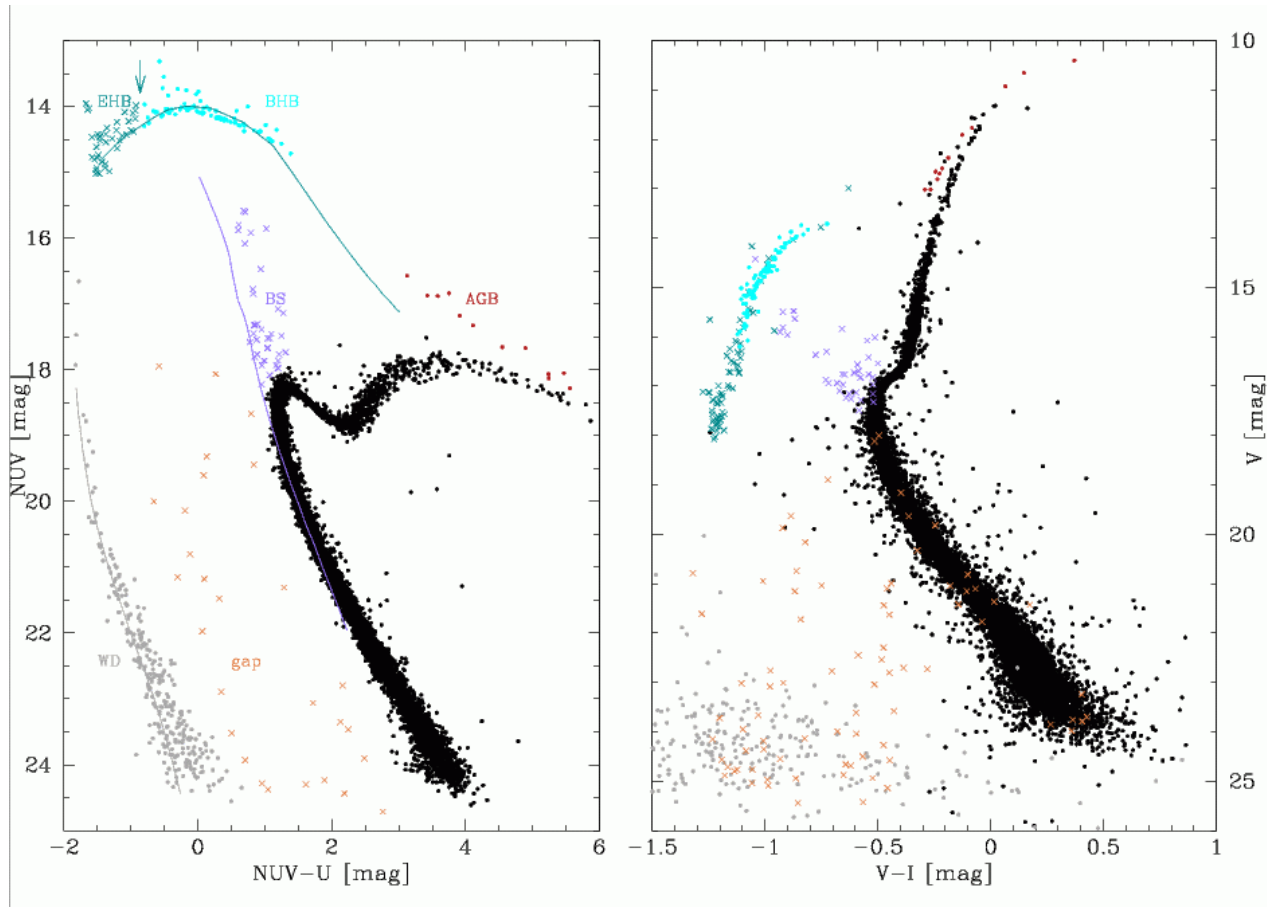}
\caption[fig_cmd_pops]{Left panel: NUV~-~U CMD of NGC\,6752. EHB stars are plotted as dark green crosses, BHB stars are plotted in cyan, BSs in purple, AGB stars in red, WDs in grey, and gap sources (including CV candidates) are orange crosses. The remaining (black) sources are MS, SGB and RGB stars. For reference, we also include a theoretical zero-age horizontal branch (ZAHB, dark green line) and zero-age main sequence (ZAMS, purple line), and a WD cooling sequence (grey line). Right panel: Optical CMD of NGC\,6752, using data from the ACS Survey Catalogue. Counterparts to NUV/U sources are plotted with the same colours and symbols as in the left panel. Optical sources with no NUV~-~U counterpart are categorised according to their V~-~I CMD position and coloured accordingly. See text for details.}
\label{fig_cmd_pops}
\end{figure*}

In order to match the catalogues from the different wave-bands, we first transformed all of the catalogues to the WCS of the master WFC3/NUV image. This has one of the largest fields of view of our images, but is not so crowded that locating matching sources becomes problematic. For each catalogue, we created a reference list of up to 16 sources (not all sources were visible in all images) which are easily visible in the image to be transformed and in the NUV reference image. We then used the {\tt IRAF} tasks {\tt geomap} and {\tt geoxytran} to determine the transformation required to shift the catalogue to the NUV WCS, allowing for shifts in the x and y directions, scale changes, and rotation. This gave a reasonable fit to the new WCS. We repeated the process using up to 48 reference sources to refine the transformation. The (rms) residual errors in the transformation were small: 0.28\,pixels (7\,milli-arcsecond (mas)) error in the FUV transformation, and a maximum of 0.11\,pixels (4.4\,mas) in the WFC3 and ACS catalogue transformations.

We searched for matches between all of the catalogues (noting that the ACS catalogue from Sarajedini et al. contains only sources that are visible in both the V- and I-band), allowing for a 2 STIS pixel ($0\farcs05$) difference in the STIS and WFC3 positions, and a 1 WFC3 pixel ($0\farcs04$) difference between the various WFC3 catalogues and between WFC3 and ACS positions. Table \ref{matchtable} gives the number of matches found in each case. Note that only V- and I-band sources with a match in at least one of our WFC3 datasets are included. This gives a total catalogue of 39411 sources. The full catalogue is available in the online journal, and a portion of it is shown in Table \ref{cattable}.

Table \ref{matchtable} also gives the percentage and number of these matches that we expect to be spurious matches. The number of false matches expected when matching two catalogues depends on the number of matches found, the matching tolerance (i.e. the area of the image `taken up' by a single source), and the number of sources in each catalogue. When matching the FUV and NUV images, we calculated this value for the entire FUV field of view; for all other matches we used a circle centred on the cluster core (see Section \ref{clustercentre}) to represent the overlapping region. As the area used includes the core, where we expect to find the most spurious matches, the numbers quoted in Table \ref{matchtable} can be considered as upper limits on the expected percentage and number of false matches. We note that the method used, described more fully in Knigge et al. (2002), does not take into account the increase in stellar density towards the cluster centre.

\subsection{Improving the Astrometry}
\label{ucac3}

The standard world coordinate system (WCS) provided with \textit{HST} data is based on the original guide star catalogue (GSGC\,1), with absolute position accuracy of $1-2\arcsec$. In order to compare our results to those in the literature (e.g. the position of the centre), it was necessary to improve the accuracy of the absolute astrometry. To do this, we used the third U.S. Naval Observatory CCD Astrograph Catalog (UCAC3; Zacharias et al. 2009). The astrometry provided in UCAC3 is on the Hipparcos (or Tycho) system and has astrometric errors of $15-20$\,mas for stars in the $10-14$\,mag range in V- and R-band.

We located 23 stars from the UCAC3 catalogue that could also be identified in our catalogue, and updated the astrometric solution for the WCS for each of our images. As in Section \ref{matching}, we repeated this process using a further 70 sources, to get a more precise transformation. The rms error between positions in our catalogue and the UCAC3 sample was $\approx0.15\arcsec$.

\section{The Colour-Magnitude Diagram}
\label{cmd}

The NUV~-~U and V~-~I CMDs for NGC\,6752 are shown in Figure \ref{fig_cmd_pops}. Different stellar populations are highlighted: WD candidates are shown in grey; BSs in purple; asymptotic giant branch (AGB) stars in dark red and `gap' sources as orange crosses. These objects are located between the MS and WD sequence and are where we would expect to find MS-WD binaries, whether they are interacting (CVs) or non-interacting binaries.

The sources marked in black are MS, sub-giant branch (SGB) or red-giant branch (RGB) stars. Where a source can be clearly identified as belonging to a certain stellar population in the NUV~-~U CMD, it is also marked as being of that population in the V~-~I CMD. The location of sources on the V~-~I CMD agrees well with the expected position based on the NUV~-~U CMD. Sources that are not in the NUV~-~U CMD are then classified according to their position in the V~-~I CMD. This method results in a handful of sources that appear to lie on the MS in the V~-~I CMD, but are marked as gap or WD sources; they clearly belong to that category in the NUV~-~U CMD. The spurious CMD positions of these sources mean that they may be false matches, or may be binary systems in which the redder source dominates at redder wavelengths, while the blue sources dominates in the bluest bands. There are also some sources that appear to be AGB stars from the V~-~I CMD that were categorised as RGB stars in the NUV~-~U catalogue. For consistency, we retain the NUV~-~U classification where one exists.

The horizontal branch (HB) is divided into blue horizontal branch (BHB) stars (light cyan), and extended or extreme horizontal branch (EHB) stars (dark green crosses). We define EHB stars to be HB stars which are bluer than an apparent gap in the NUV~-~U HB corresponding to around 16,500K (marked with an arrow in Figure \ref{fig_cmd_pops}). This is consistent with the usual definitions for EHB stars (e.g. Momany et al. 2004, Brown et al. 2010).

Figure \ref{fig_cmd_pops} also shows a theoretical zero-age horizontal branch (ZAHB, dark green line), zero-age main sequence (ZAMS, purple line), and a WD cooling sequence (grey line). These were created using a distance of $4.0$\,kpc, reddening of $E(B-V)=0.04$\,mag (Harris 1996, 2010 edition) and metallicity of $[Fe/H]\simeq-1.5$ (Gratton et al. 2005).

\begin{figure}
\includegraphics[width=0.5\textwidth]{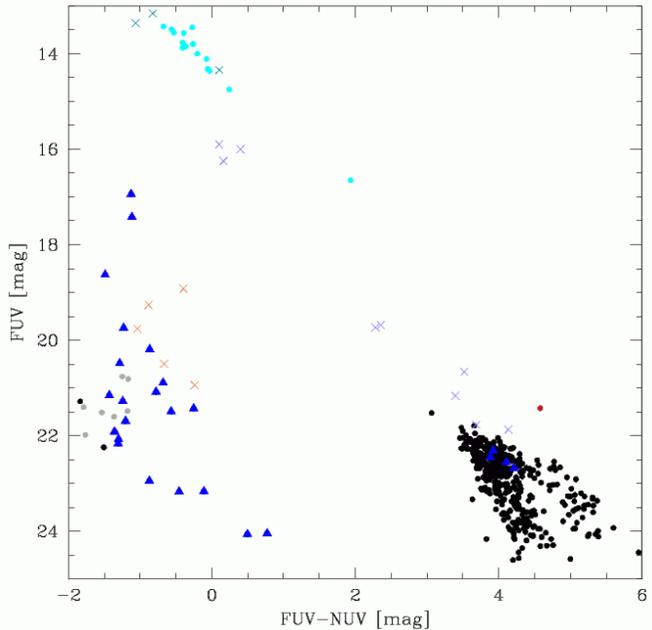}
\caption[fig_cmd_fuv]{FUV~-~NUV CMD of NGC\,6752. Sources are coloured according to their position in the NUV~-~U CMD (see Figure \ref{fig_cmd_pops} and text). EHB stars are plotted in dark green, BHB stars in cyan, BSs in purple, AGB stars in red, WDs in grey, and gap sources (including CV candidates) are in orange. Blue triangles indicate sources that have no counterpart in the U-, B-, V- or I-bands. The remaining (black) sources are MS, SGB and RGB stars.}
\label{fig_cmd_fuv}
\end{figure}

The FUV~-NUV CMD is shown in Figure \ref{fig_cmd_fuv}. The colours are as per Figure \ref{fig_cmd_pops}. The blue triangles indicate sources for which there is no U-, V- or I-band counterpart. The FUV~-~NUV CMD is not populated enough to clearly distinguish between WD and gap sources.

\section{Identification of X-Ray Counterparts}
\label{xray}

\begin{figure*}
\includegraphics[height=5cm]{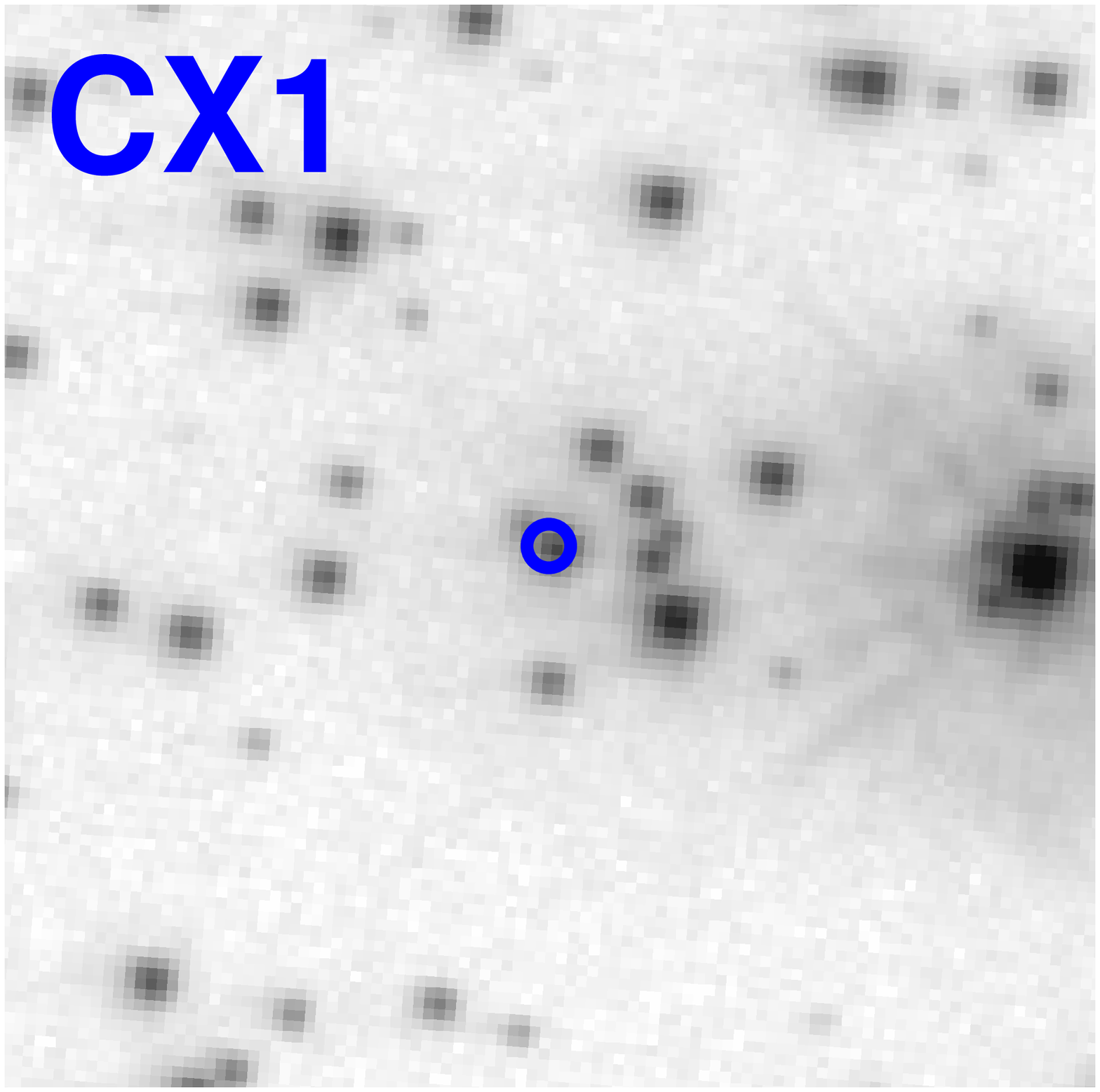}
\includegraphics[height=5cm]{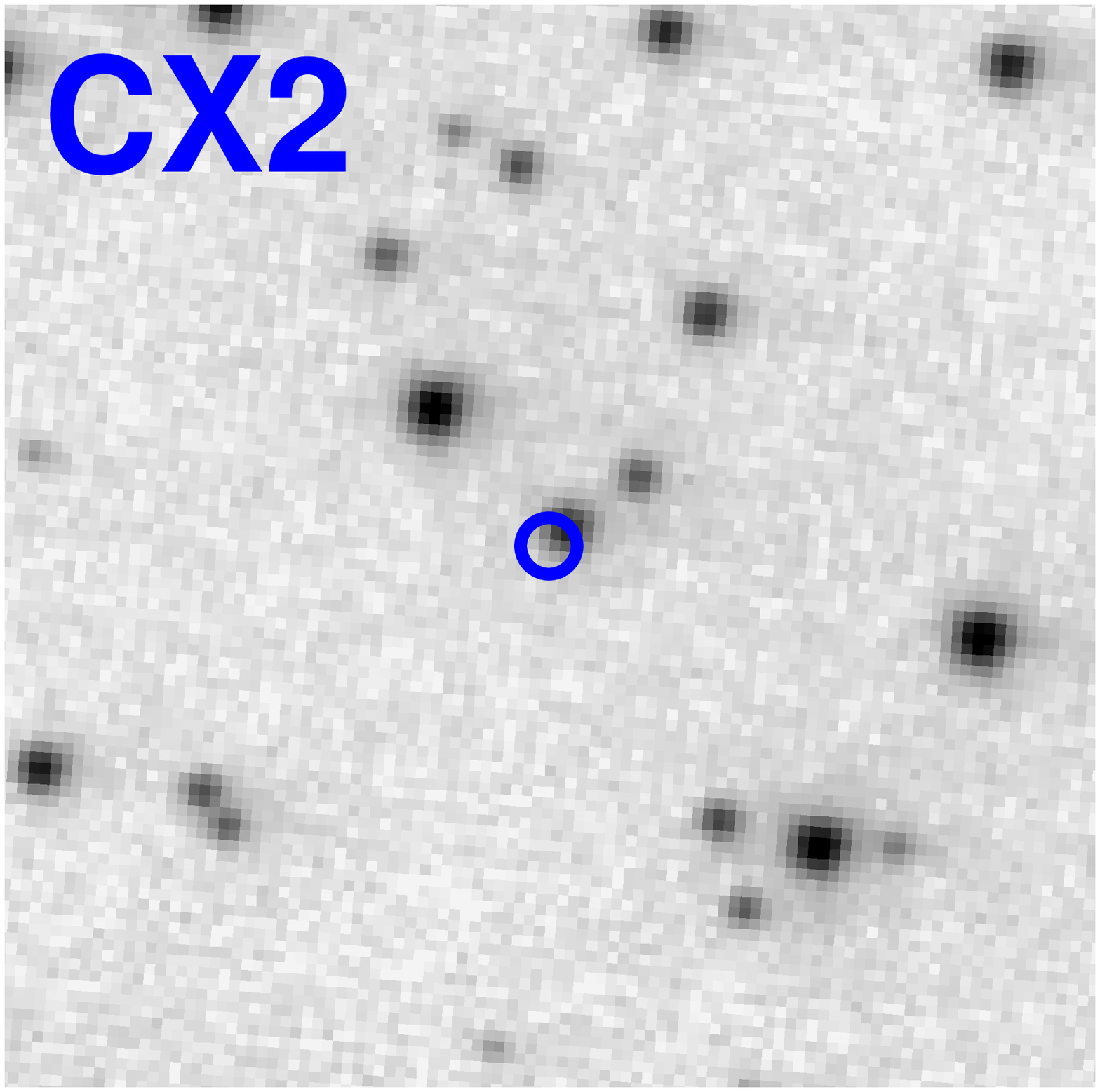}
\includegraphics[height=5cm]{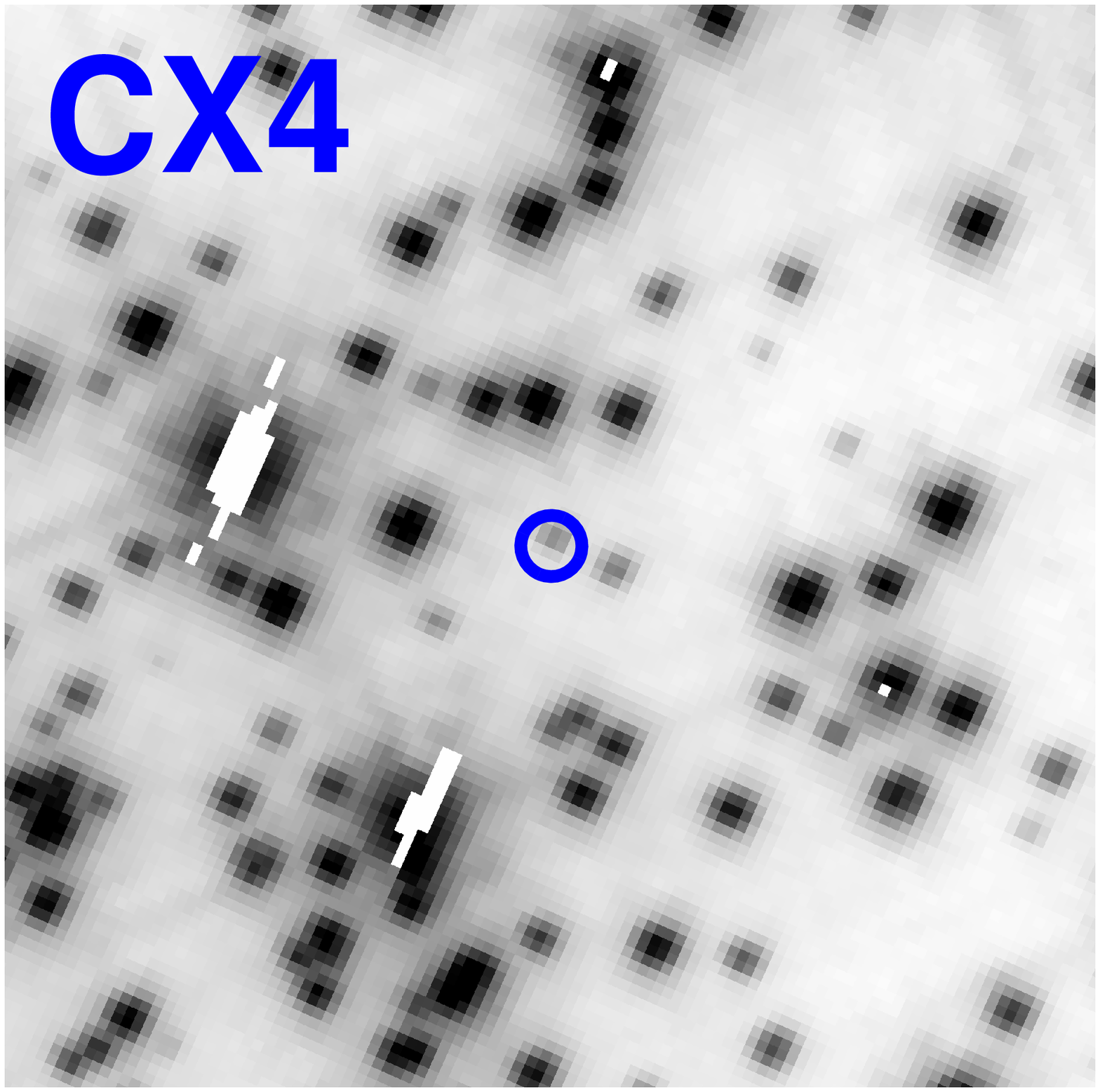}\\
\includegraphics[height=5cm]{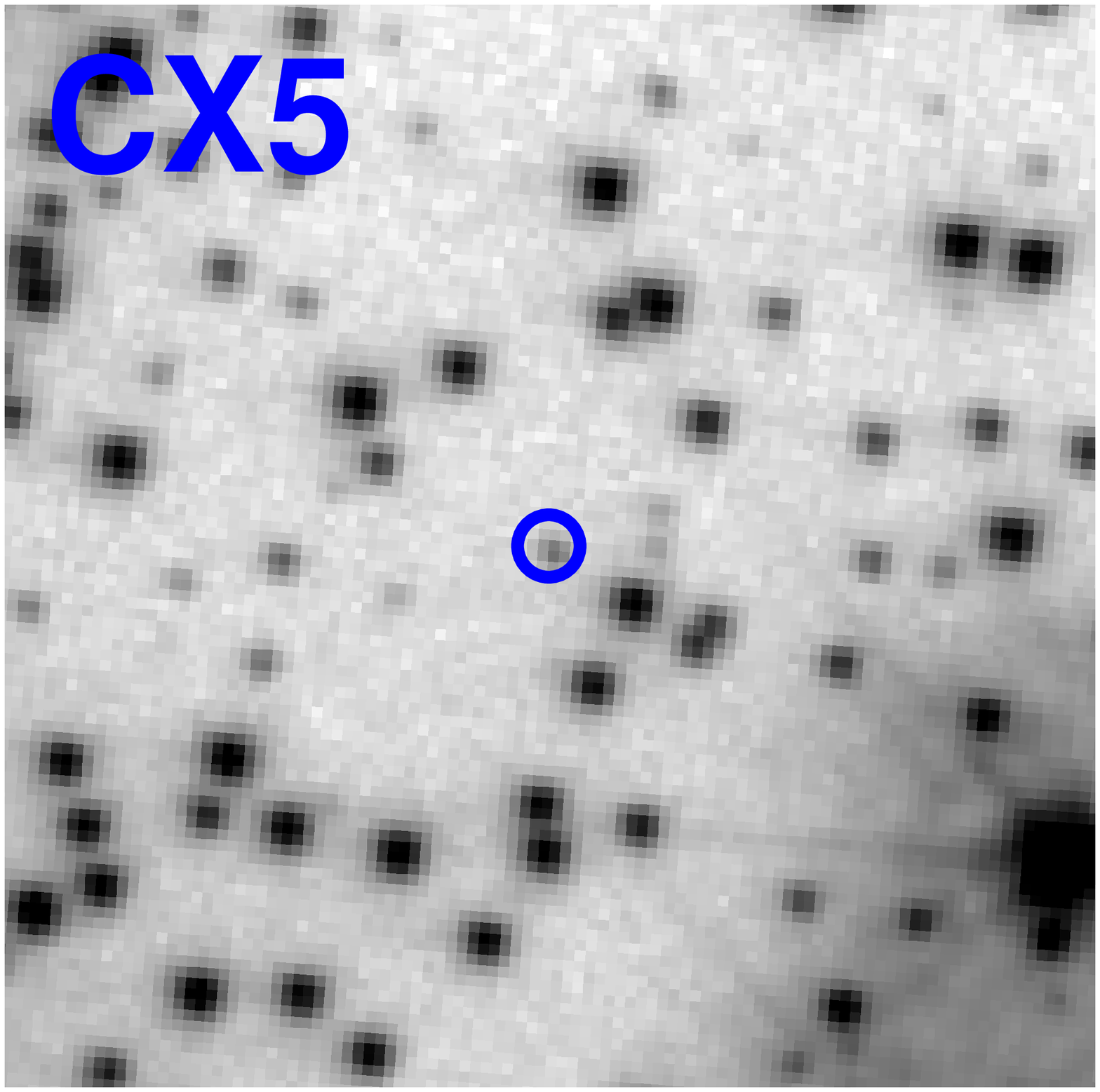}
\includegraphics[height=5cm]{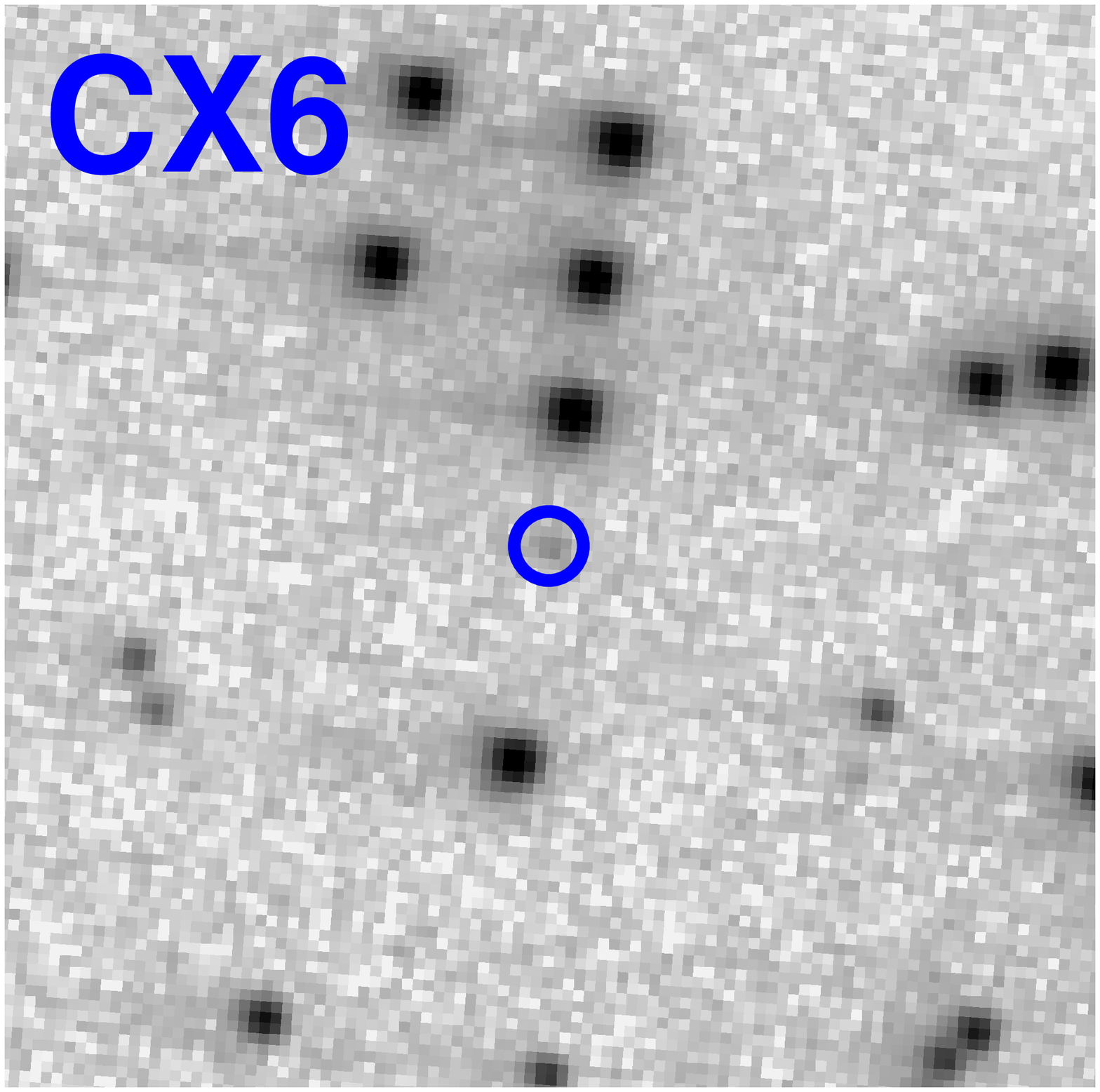}
\includegraphics[height=5cm]{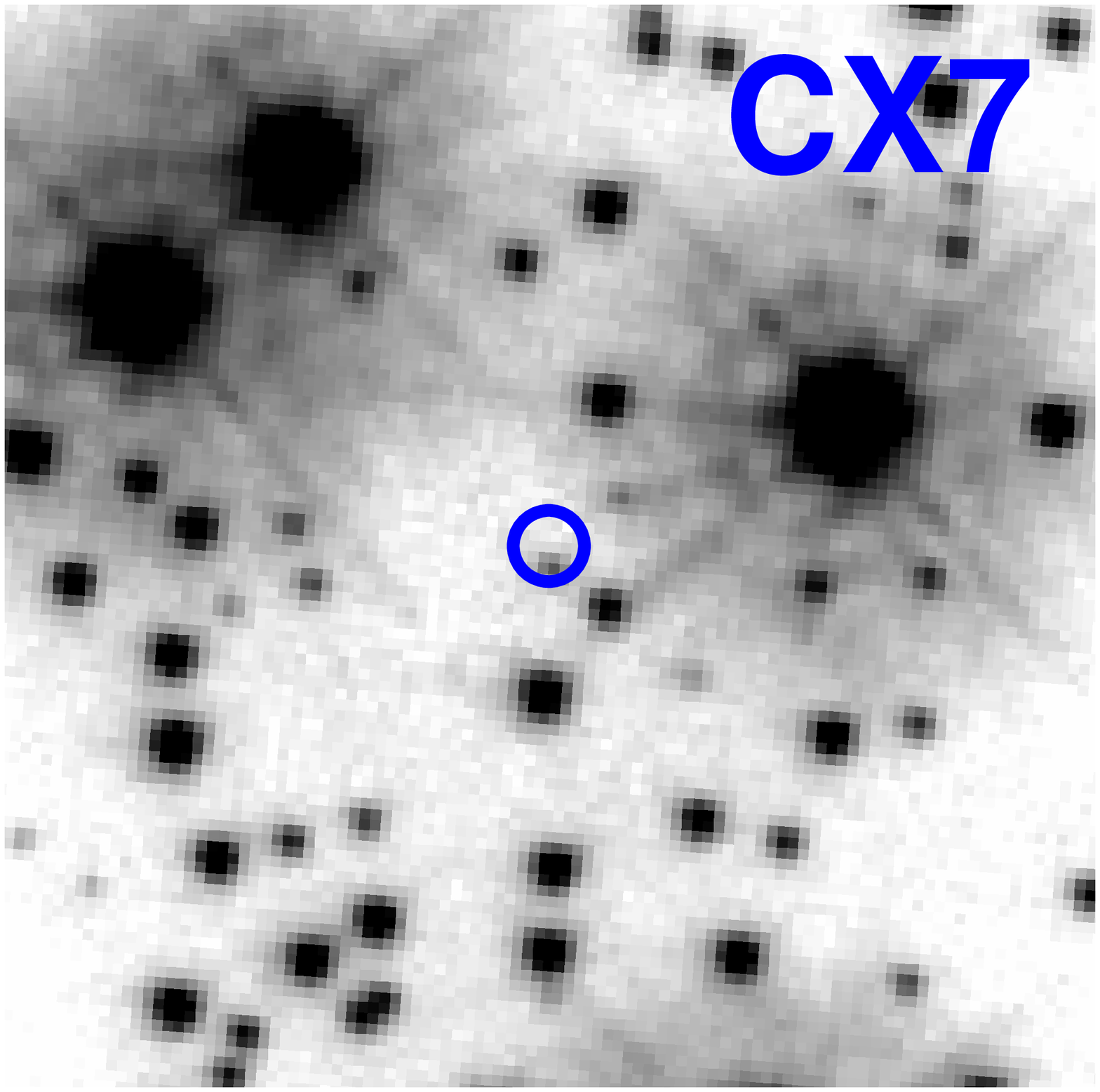}\\
\includegraphics[height=5cm]{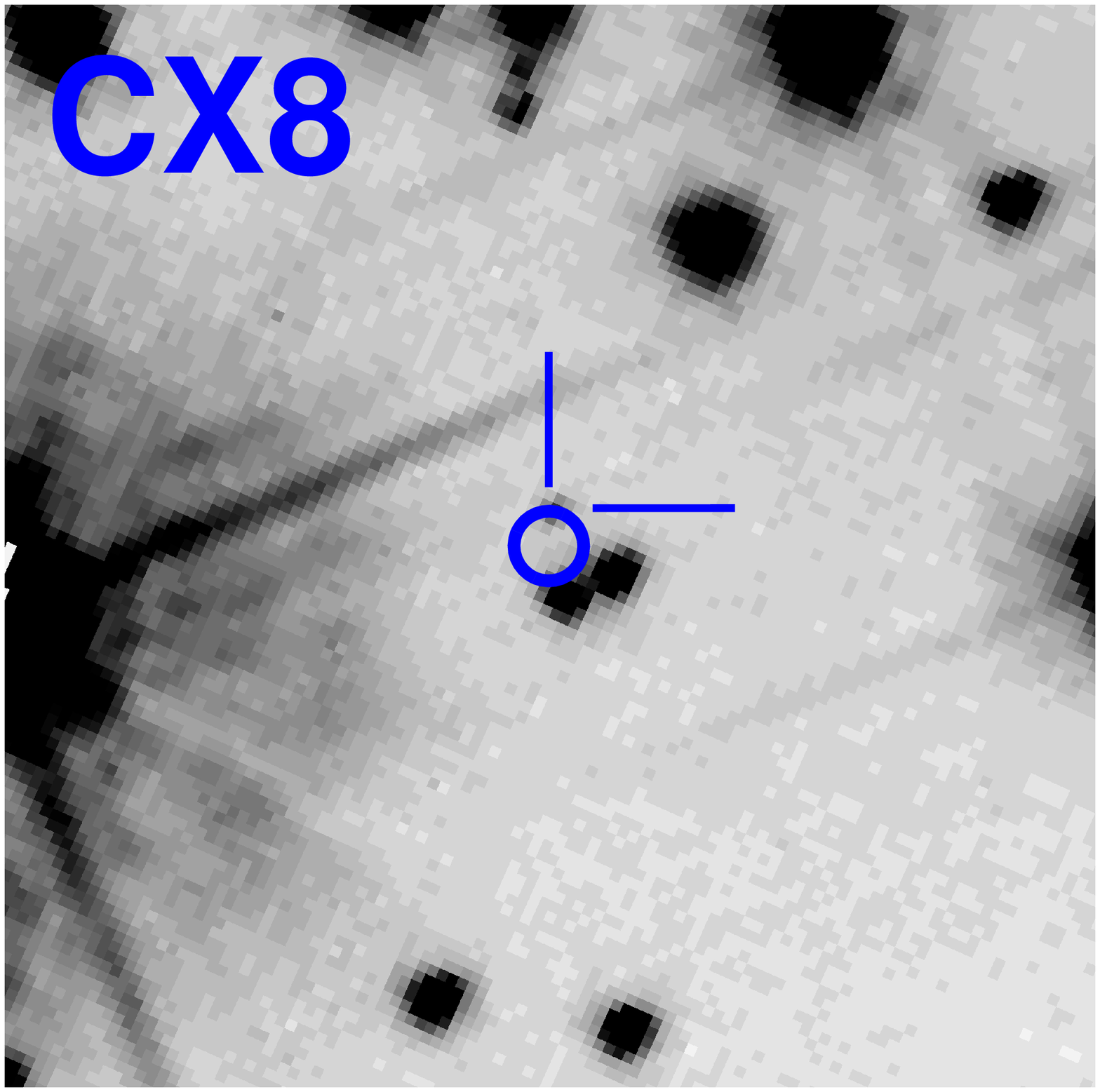}
\includegraphics[height=5cm]{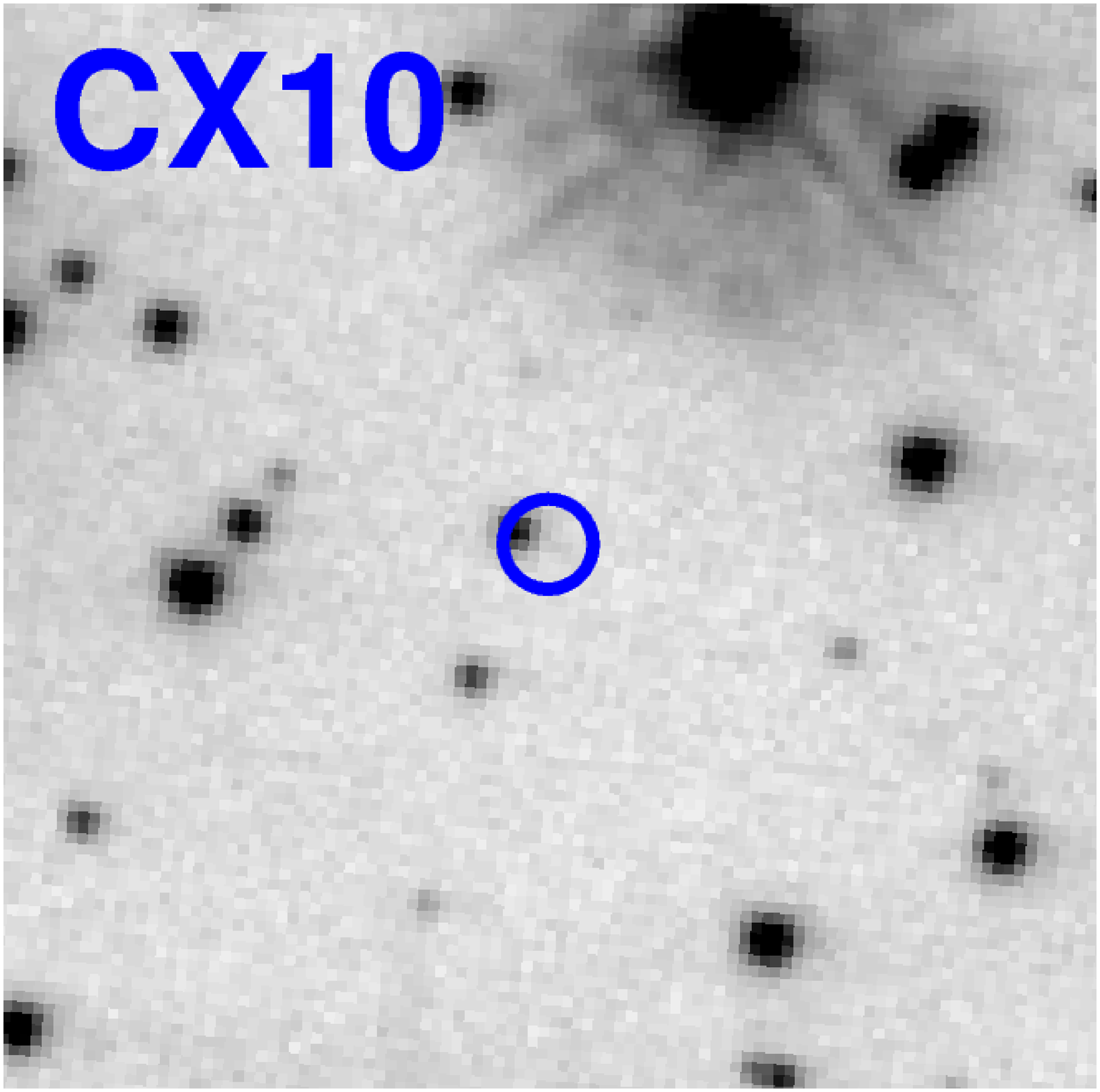}
\includegraphics[height=5cm]{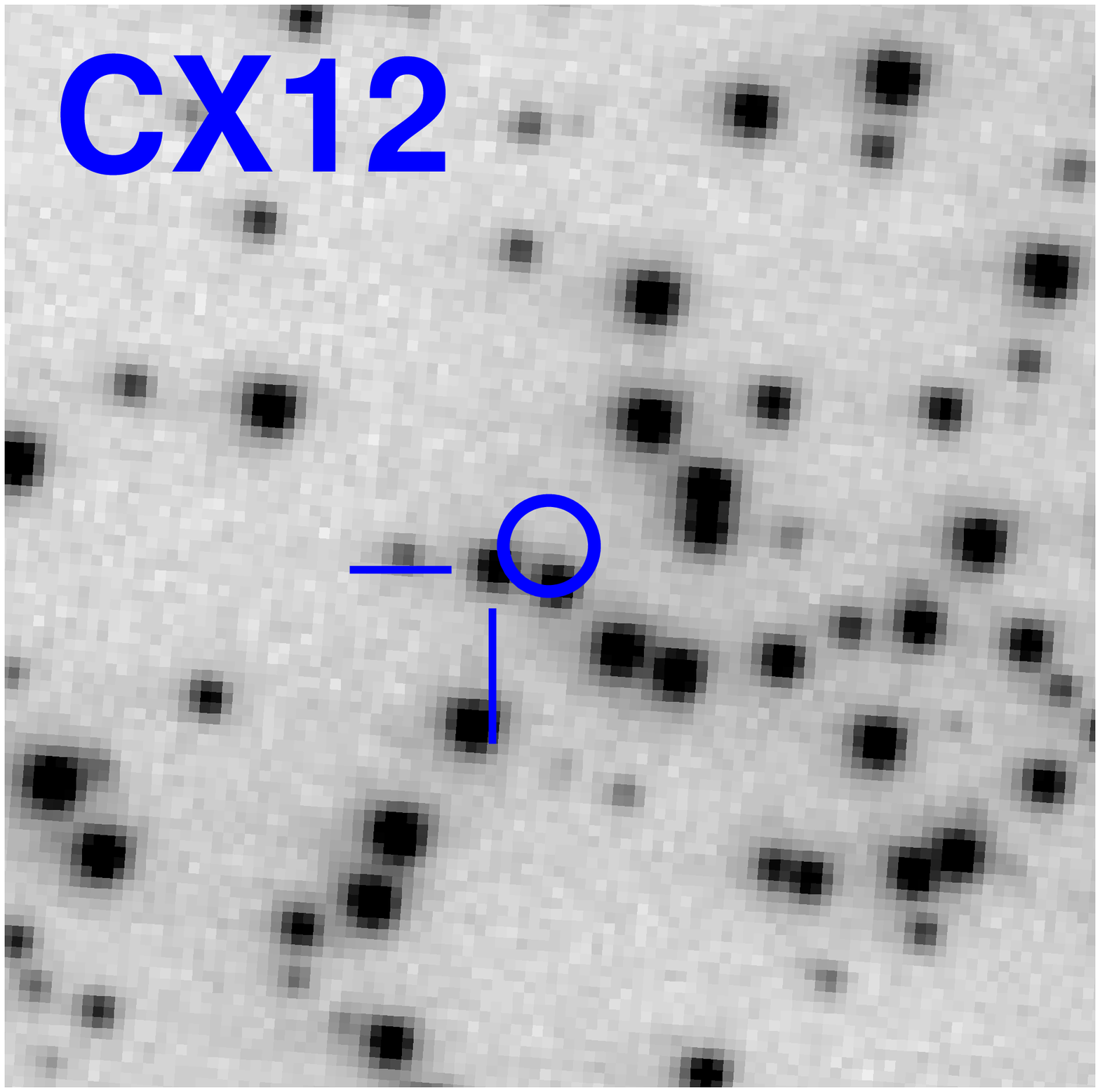}\\
\includegraphics[height=5cm]{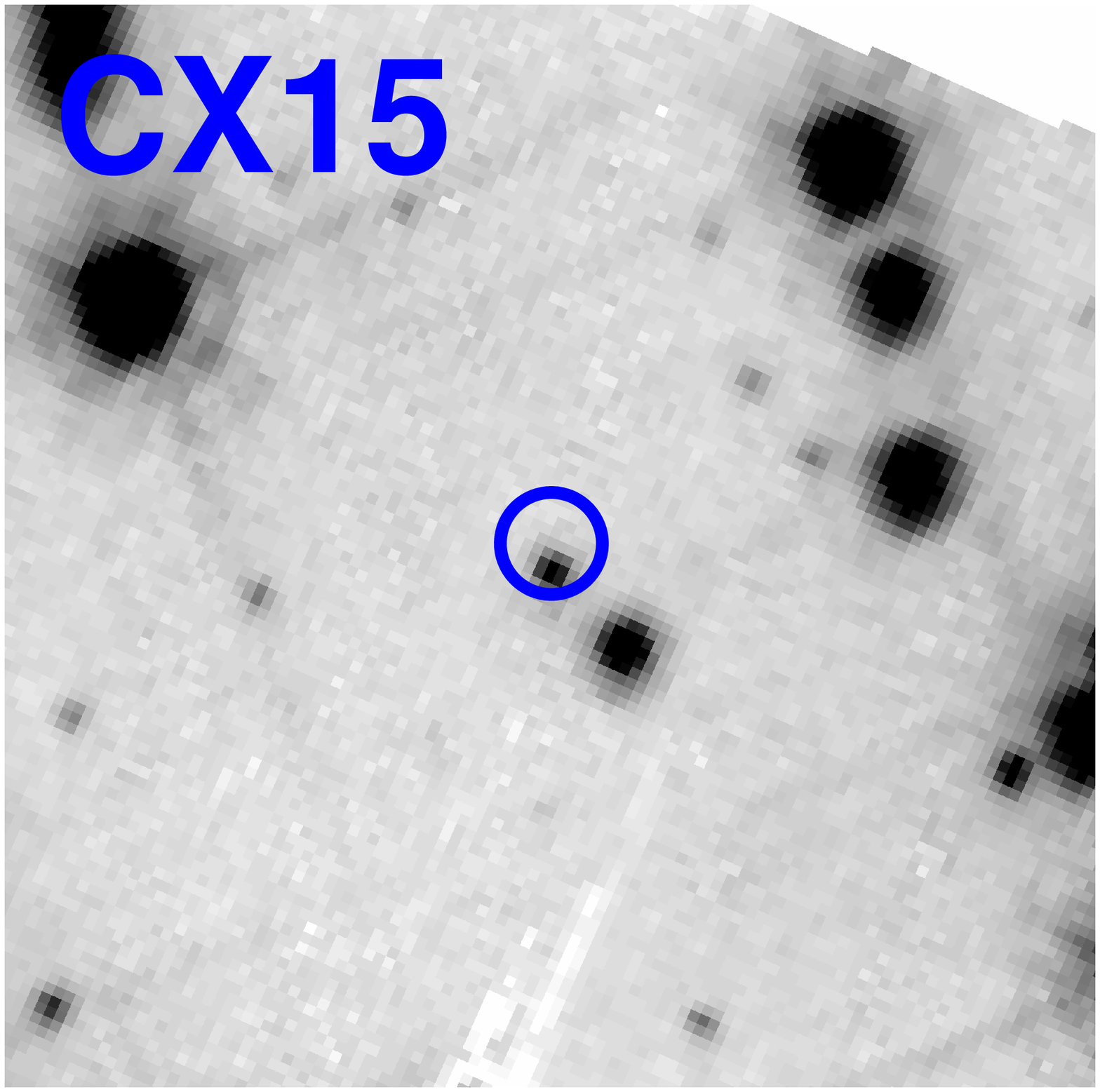}
\includegraphics[height=5cm]{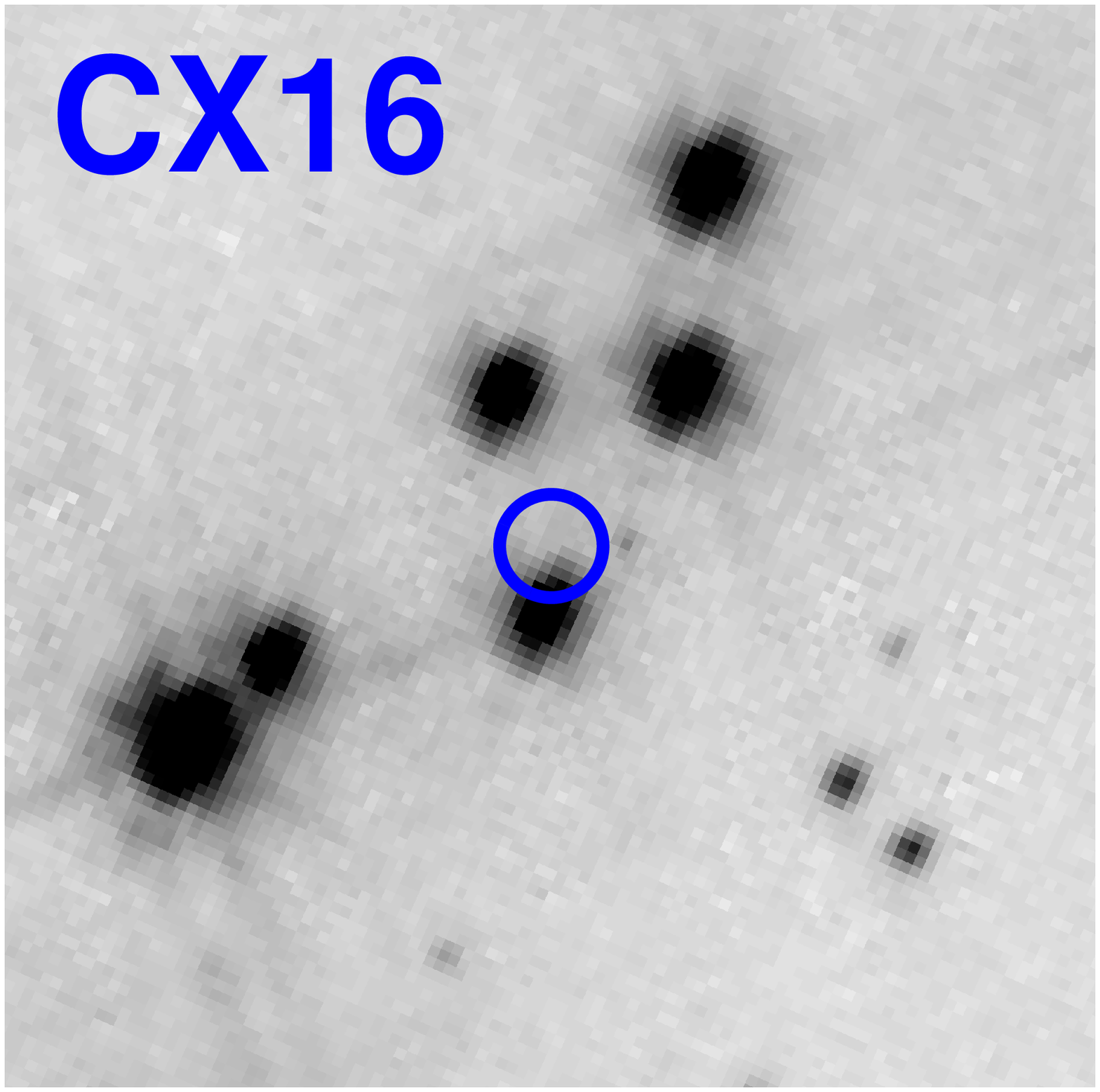}
\caption[fig_finder]{Finder charts showing possible counterparts to X-ray sources. In each case, the field covers $4\arcsec \times 4\arcsec$ and North is up and East is to the left. The circles show the 90\% confidence level uncertainty in X-ray position that we searched for counterparts. For sources CX\,1, 2, 5, 6, 7, 10 and 12, the image is the NUV `master' image created by combining 15 individual F225W exposures. For sources CX\,4, 8, 15 and 16, the image is a combination of 6 U-band images taken with the F390W filter. The grey-scale is varied to enhance the visibility of the counterparts. Sources CX\,3, 9, 13, 17 and 19 are outside our WFC3 field of view. No sources were detected within the 90\% confidence error circle of CX\,11, 14 or 18.}
\label{fig_finder}
\end{figure*}

\begin{figure*}
\includegraphics[angle=270,width=0.9\textwidth]{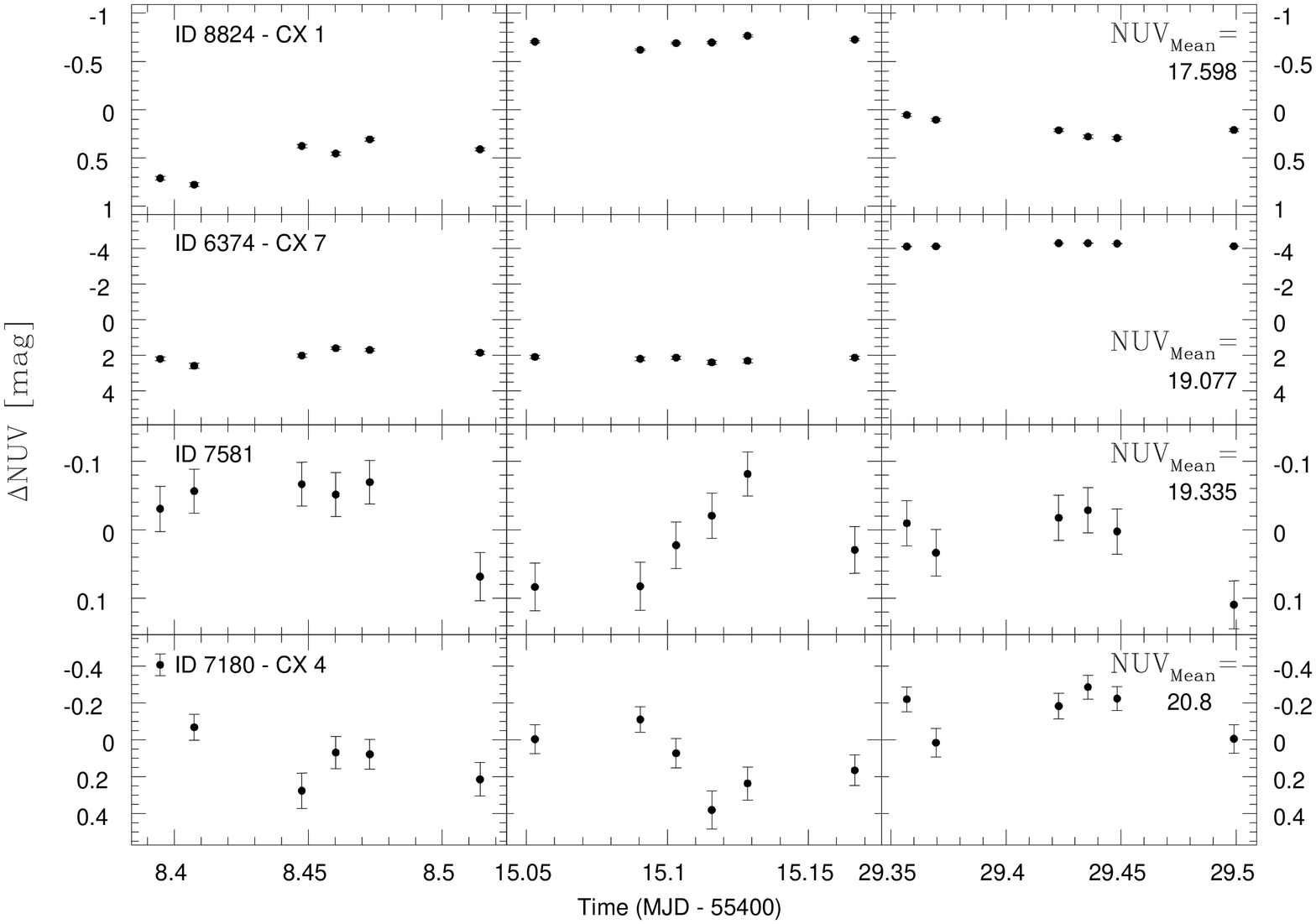}
\caption[fig_lightcurves]{Light curves of four likely optical counterparts to X-ray sources. The NUV variation from the mean NUV magnitude is shown ($\Delta NUV=NUV-NUV_{Mean}$). This value does not necessarily match the magnitude stated in the catalogue, which is an overall (weighted average) magnitude, as opposed to a simple mean value. Top panel: Source 8824, the optical counterpart to CX\,1, which shows a brightening in the middle epoch and is a likely DN. Second panel: Source 6374, the counterpart to CX\,7. This source is $\approx6$\,mag brighter in the third epoch than the first two. We conclude that this source is a DN. Third panel: Source 7581, which may be the previously unidentified counterpart to CX\,12 and is an SX~Phoenicis star. Bottom panel: Source 7180, which corresponds to X-ray source CX\,4 which is a known DN.  See text for details.}
\label{fig_lightcurves}
\end{figure*}

Based on $Chandra$ observations, Pooley et al. (2002) identified 19 X-ray sources within the $115\arcsec$ half-mass radius of NGC\,6752, including six within the $10\farcs5$ core radius, down to a limiting luminosity of $L_{X}\approx10^{30}\,ergs\,s^{-1}$. Using archival \textit{HST} data, Pooley et al. suggested 12 optical counterparts. These include 10 CV candidates, one to three RS CVn or BY Dra sources (based on their X-ray-to-optical flux ratio limits), and one or two that are background objects. Seven of the X-ray sources had no detectable optical counterparts. One of the CVs, CX\,4, is now known to be a DN (Kaluzny \& Thompson 2009).

Sixteen of the $Chandra$ X-ray sources are in the field of view of our WFC3 observations, and 7 are also in the field of the FUV image. Using the most obvious known matches (the DN and other bright CVs), we made a first attempt to register the X-ray source positions to our (Tycho based) WCS. Comparing the positions of all X-ray sources to our catalogue revealed more matches that could be considered `safe', including two further DNe (see Section \ref{xray_dn}). We used the (now three) DN counterparts to refine the correction. We found that the $Chandra$ positions quoted by Pooley et al. (2002) should be shifted by $0\farcs540$ in right ascension and $-0\farcs055$ in declination in order to best match the positions of the DNe, giving an rms offset between the X-ray and optical positions of the DNe of $0\farcs012$ in right ascension and $0\farcs003$ in declination.

Using the empirical equation derived by Kim et al. (2007) for the positional uncertainty of $Chandra$ X-ray sources, and ignoring the off-axis angle in the $Chandra$ observations (which is likely to be negligible given the relatively small field of view), we reconstructed the 90\% confidence level uncertainty of the X-ray source positions. This was the dominant source of error in the X-ray positions; for all but the very brightest sources, the $Chandra$ uncertainty was several times larger than the estimated error on our boresight correction.

We compared the positions of all X-ray sources to our catalogue to search for matches, using the 90\% confidence level uncertainty in X-ray position as the maximum matching radius. Figure \ref{fig_x_all} shows the FUV and NUV images with the X-ray source positions overplotted, and Figure \ref{fig_finder} presents finder charts of all counterparts.

Using the method described in Section \ref{matching}, we calculate that we expect to find 2 spurious matches to the 16 X-ray sources in our field of view. As outlined in the following sections, we find 12 sources within the regions searched, at least 7 of which are likely to be the real optical counterpart, based on their CMD positions or light curves. For one X-ray source (with a likely `real' counterpart), there are 2 sources within the maximum matching radius; clearly, at least one of these `matches' is spurious. Of the remaining 5 matches, which are MSs or were only detected in one wave-band, the calculation suggests that 1 is likely to be a spurious match, while the other 4 may be real counterparts. We caution that Poisson errors on these numbers mean that they are approximations only.

We searched for hints of periodic variability in the light curves of every potential X-ray counterpart. Following Thomson et al. (2010), photometry was performed on the individual FUV exposures, using the overall FUV catalogue as input to {\tt daophot} (see Section \ref{phot}). As described in Section \ref{phot}, photometry was carried out on the NUV and U-band data using DOLPHOT, which gives individual magnitudes for each source in each of the 18 individual images as part of its output. For each source, we calculated a reduced $\chi^{2}$ value by comparing each magnitude measurement to the mean magnitude for the source. We then compared the $\chi^{2}$ value of each X-ray counterpart candidate to other sources of similar magnitude. Outliers found in this way are likely to be variable. We produced a power spectrum for each of these outliers and used a randomisation test to estimate a false alarm probability (FAP); this is the probability that a higher peak could be produced at any frequency if the positions of the data were shuffled. Table \ref{xraytable} summarises the results. Figure \ref{fig_lightcurves} shows the light curves of the four most interesting counterparts.

\begin{figure*}
\includegraphics[angle=270,width=0.95\textwidth]{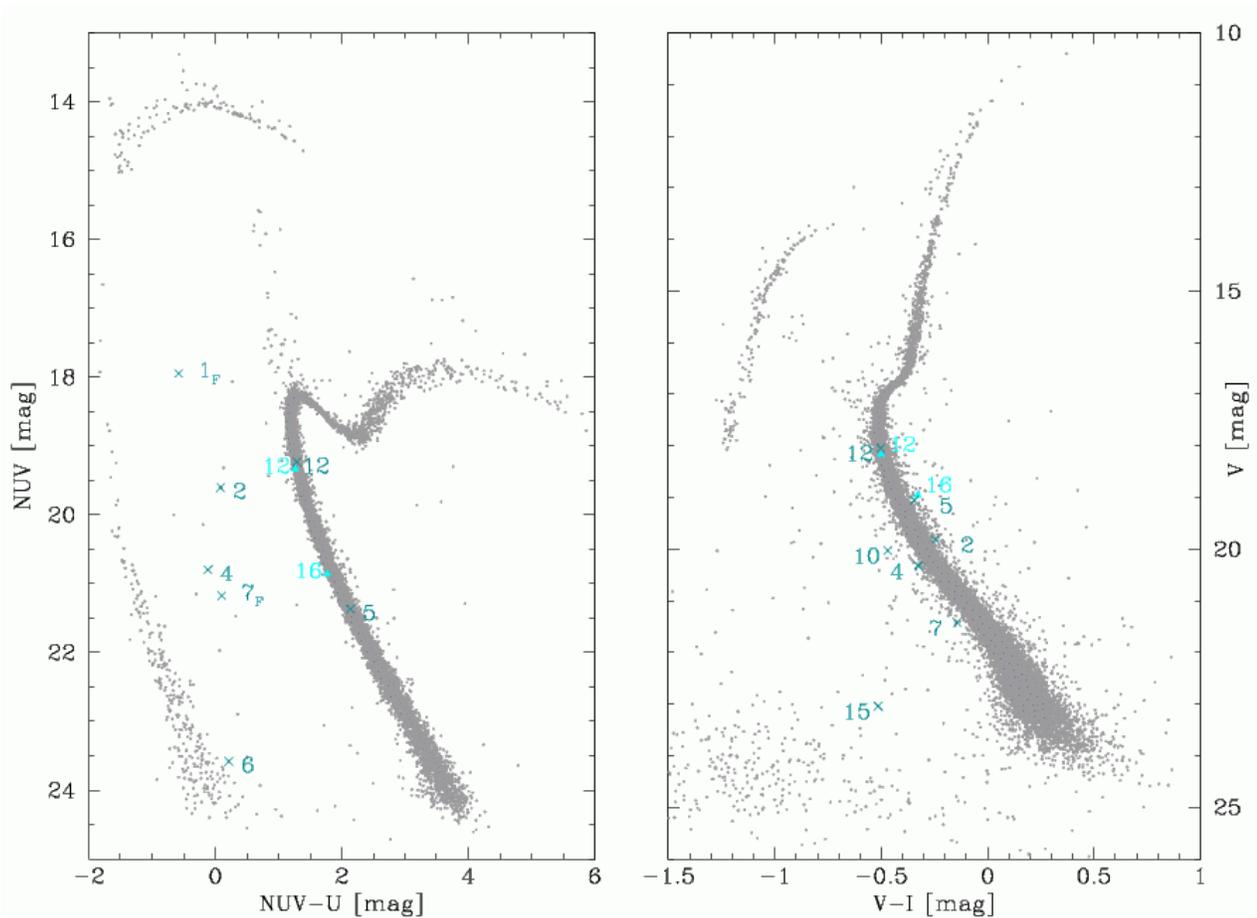}
\caption[fig_cmd_var]{NUV~-~U (left panel) and V~-~I (right panel) CMD of NGC\,6752. The most likely counterpart to the X-ray sources are shown as dark crosses and labelled with their X-ray source ID. The most likely counterparts to CX\,1 and CX\,7 exhibit strong variation in their light curves (see Section \ref{xray_dn}). The NUV magnitude plotted is the average magnitude of the source when in its `fainter state' and is marked with a subscript `F'. Interesting optical sources that lie just outside the $Chandra$ error circle of CV\,12 and CX\,16 are marked with cyan triangles and the X-ray source ID. See Section \ref{xray} for details of X-ray matching.}
\label{fig_cmd_var}
\end{figure*}

Our main result is that two of the X-ray sources that Pooley identified as CV candidates are, in fact, dwarf novae (DNe). This confirms the CV nature of these sources, and determines their sub-class. For two X-ray sources, we identified new optical counterpart candidates, including one which is a variable. In the following subsections we provide details of these and all other potential counterparts to each of the X-ray sources in our field of view. The CMD positions of all of the counterparts discussed (where CMD information is available) are shown in Figure \ref{fig_cmd_var}.

\subsection{CX\,1 and CX\,7: Two Dwarf Novae}
\label{xray_dn}

Very few DNe have been found in GCs (Pietrukowicz et al. 2008), although to what extent this is due to selection effects remains unclear (Knigge 2011). Prior to this study, only one DN was known in NGC\,6752 (Kaluzny \& Thompson, 2009). Two X-ray sources which were previously known to be CVs, CX\,1 and CX\,7, show DN-like outbursts in our data. This takes the number of known DNe in NGC\,6752 up to three, more than any other cluster.

Pooley et al. (2002) identified source CX\,1 as a CV. This source matches with number 8824 in our catalogue, which was outside the FUV field of view, but was detected in our NUV and U-band images. The light curve from the NUV images, shown in Figure \ref{fig_lightcurves}, shows that the source is $\approx1.5$\,mag brighter in the middle observing epoch than in epochs one and three. Subtracting the mean magnitude from each epoch, a Lomb-Scargle power spectrum shows that the small scale NUV variability could be fit with a period of 4.1\,hours, but the peak in the power spectrum is only marginally significant. Figure \ref{fig_lightcurves} also shows that the amplitude of the variation is suppressed somewhat during the outburst. This source exhibits X-ray emission, has short time scale variability, as well as an outburst, and is situated in the gap between WD and MS on the CMD. We therefore argue that this source should be considered to be a confirmed CV of the DN sub-class.

CX\,7 was first identified as a CV candidate by Bailyn et al. (1996), who found a period of 3.7\,hours. This source corresponds to source 6374 in our catalogue and was identified in all of the wave-bands used. The second panel of Figure \ref{fig_lightcurves} presents the light curve from the NUV data, which clearly shows a 6 magnitude outburst in the final observing epoch. This indicates that this source is a DN. Subtracting the mean magnitude measured in each observing epoch from the NUV data we found a tentative best-fit period of 3.5\,hours, which is not quite consistent with the result of Bailyn et al., but, again, the data are not sufficient to draw strong conclusions regarding the period.

\subsection{CX\,8, CX\,12 and CX\,16: New Optical Counterpart Candidates}
\label{xray_new}

Two of the X-ray sources without optical counterparts in Pooley et al. (2002), CX\,8 and CX\,12, have new potential optical counterparts in our data. For a further source for which Pooley et al. (2002) did suggest a counterpart, we have found a source that may be a better match.

CX\,8's error circle contains source ID 24678. CX\,8 is outside the FUV field of view, but 24678 was detected in U- and B-bands. The light curve does not exhibit periodic variability, but the CMD positions indicate that it is a faint gap source. We suggest that this source may be a CV.

\begin{landscape}
\begin{table}
\caption{Properties of potential counterparts to $Chandra$ X-ray sources.}
\label{xraytable}
\begin{tabular}{cccccccccccccll}
\hline
\hline
1        & 2          & 3            & 4            & 5                & 6     & 7     & 8         & 9      & 10     & 11     & 12    & 13                & 14   \\
\hline
ID$_{X}$ & ID$_{cat}$ & R.A.       & Decl.       & L$_{X,0.5-2.5}$  &FUV   &NUV     &$\sigma$NUV &U     &B     &V      &I       & Previous Status         & CMD position,\\
         &            & (hh:mm:ss) & (deg:mm:ss) & (ergs\,s$^{-1}$) &(mag) &(mag)   &(mag)       &(mag) &(mag) &(mag)  &(mag)   &                         & Comments \\
\hline
CX\,1 & 8824   & 19:10:51.134 & -59:59:11.83 & $2.1\times10^{32}$ & ...    & 17.700 & 0.539 & 18.528 & ...    & ...    & ...    & CV$^{P}$                & gap, DN\\
CX\,2 & 12078  & 19:10:56.009 & -59:59:37.38 & $6.0\times10^{31}$ & ...    & 19.604 & 0.423 & 19.520 & 19.728 & 19.818 & 20.064 & CV$^{P}$                &  gap, (CV)\\
CX\,3 & ...    & 19:10:40.354 & -59:58:41.34 & $5.3\times10^{31}$ & ...    & ...    & ...   & ...    & ...    & ...    & ...    & CV$^{P}$                & ...\\
CX\,4 & 7180   & 19:10:51.583 & -59:59:01.76 & $4.0\times10^{31}$ & 19.756 & 20.800 & 0.212 & 20.914 & 21.002 & 20.328 & 20.653 & CV$^{P}$, DN$^{K}$      & gap, CV\\
CX\,5 & 7796   & 19:10:51.410 & -59:59:05.16 & $3.6\times10^{31}$ & ...    & 21.371 & 0.140 & 19.234 & 19.089 & 19.057 & 19.402 & CV/BY Dra$^{P}$         & MS, (CV)\\
CX\,6 & 10808  & 19:10:51.499 & -59:59:27.05 & $2.2\times10^{31}$ & ...    & 23.583 & 0.652 & 23.367 & ...    & ...    & ...    & CV$^{P}$                & WD, (CV)\\
      & 34433  &              &              &                    & ...    & ...    & ...   & ...    & 23.102 & ...    &  ...   &                         &  ...\\
CX\,7 & 6374   & 19:10:51.504 & -59:58:56.77 & $1.9\times10^{31}$ & 20.941 & 20.412 & 3.067 & 21.084 & 21.650 & 21.424 & 21.567 & CV$^{P}$                & gap, DN\\
CX\,8 & 24678  & 19:11:02.981 & -59:59:41.94 & $2.1\times10^{31}$ & ...    & ...    & ...   & 24.011 & 23.648 & ...    & ...    & ...                     & gap\\
CX\,9 & ...    & 19:10:51.756 & -59:58:59.21 & $1.3\times10^{31}$ & ...    & ...    & ...   & ...    & ...    & ...    & ...    & ...                     & ...\\
CX\,10 & 9343  & 19:10:54.742 & -59:59:13.92 & $6.0\times10^{30}$ & ...    & 20.288 & 0.098 & ...    & ...    & 20.029 & 20.499 & CV$^{P}$                & gap, (CV)\\
CX\,11 & ...   & 19:10:52.411 & -59:59:05.61 & $6.2\times10^{30}$ & ...    & ...    & ...   & ...    & ...    & ...    & ...    & MSP$^{D}$, CV/Gal$^{P}$ & ...\\
CX\,12 & 7592  & 19:10:52.730 & -59:59:03.33 & $5.6\times10^{30}$ & 23.199 & 19.235 & 0.045 & 17.946 & 17.840 & 18.044 & 18.547 & ...                     & MS\\
       & 7581  &              &              &                    & ...    & 19.336 & 0.058 & 18.073 & 17.954 & 18.169 & 18.669 &                         & MS, Variable$^{E}$\\
CX\,13 & ...   & 19:10:40.601 & -60:00:06.12 & $4.6\times10^{30}$ & ...    & ...    & ...   & ...    & ...    & ...    & ...    & CV$^{P}$                & ...\\
CX\,14 & ...   & 19:10:52.075 & -59:59:09.18 & $4.2\times10^{30}$ & ...    & ...    & ...   & ...    & ...    & ...    & ...    & ...                     & ...\\
CX\,15 & 14708 & 19:10:55.834 & -59:57:45.58 & $3.2\times10^{30}$ & ...    & ...    & ...   & 22.583 & 23.295 & 23.044 & 23.558 & CV/Gal.$^{P}$           & gap/WD, (CV)\\
CX\,16 & 31669 & 19:10:42.509 & -59:58:42.88 & $3.0\times10^{30}$ & ...    & ...    & ...   & ...    & 23.484 & ...    & ...    &                         & ...\\
       & 4190  &              &              &                    & ...    & 20.861 & 0.135 & 19.090 & 18.935 & 18.949 & 19.278 & BY Dra$^{P}$            & MS$^{E}$\\
CX\,17 & ...   & 19:11:05.316 & -59:59:04.08 & $2.7\times10^{30}$ & ...    & ...    & ...   & ...    & ...    & ...    & ...    & MSP/Gal.$^{P}$          & ...\\
CX\,18 & ...   & 19:10:52.042 & -59:59:03.74 & $2.7\times10^{30}$ & ...    & ...    & ...   & ...    & ...    & ...    & ...    & ...                     & ...\\
CX\,19 & ...   & 19:10:55.613 & -59:59:17.60 & $2.2\times10^{30}$ & ...    & ...    & ...   & ...    & ...    & ...    & ...    & Close binary$^{K}$      & ...\\
\hline
\end{tabular}
\end{table}
\begin{scriptsize}
\noindent The first column is the $Chandra$ ID number from Pooley et al. (2002), followed by our ID number in Column 2. Columns 3-4 give the source position in R.A. and decl. (shifted to match our Tycho based WCS). Column 5 gives the $0.5-2.5$\,keV X-ray luminosity from Pooley et al. (2002). Columns 6-7 and 9-10 give the magnitudes in STMAG, and Column 8 gives an estimate of the variability amplitude, defined to be the standard deviation of the source from its mean magnitude. Columns 11 and 12 give the magnitudes in STMAG from the ACS Survey Catalogue. Column 13 gives previous suggestions of source type, and the final column gives our CMD position and categorisation of each source. Where a source is marked `(CV)', we detected the counterpart suggested by Pooley et al. (2002), but our data were not sufficient for us to draw any further conclusions regarding its nature.\\
\\
$^{P}$ Counterpart type suggested by Pooley et al. (2002). `Gal' indicates that the source may be a galaxy.\\
$^{K}$ Kaluzny and Thompson (2009) found that CX\,4 (their V\,25) is a DN, and suggest that CX\,19 is a close binary hosting a NS or a BH.\\
$^{D}$ The position of CX\,11 is consistent with the MSP PSR\,D from D'Amico et al. (2002).\\
$^{E}$ Source 7581 is outside the error circle of CX\,12, but is included because it shows variability.\\
Source 4190 is outside the error circle of CX\,16. We include it in the table because Pooley et al. (2002) concluded that this source is a BY Dra or RS CVn source based on its CMD position and H$\alpha$ emission.\\
\end{scriptsize}
\end{landscape}

The MS source 7592, is within the $Chandra$ error circle of CX\,12, but the FUV and NUV light curves showed no hint of periodic variability. Interestingly, source 7581, which is just outside CX\,12's error circle, does appear to vary (see third panel of Figure \ref{fig_lightcurves}), with a possible period of 4\,hours. This source is also on the MS. It was not detected in the FUV data, despite being in the field of view. Based on the variability, possible period and the fact that the CMD position makes a faint BS classification possible, we suggest that this is an SX~Phoenicis star. Recently, X-ray emission has been detected from Cepheid variables (Engle et al. 2009), possibly due to magnetic activity associated with pulsations, or the presence of an active binary companion. Source 7581 is marked on the finder chart (Figure \ref{fig_finder}).

Source CX\,16 was identified as a BY Draconis star by Pooley et al. (2002). We detected the optical counterpart they suggested, but it was slightly outside the 90\% confidence error circle of the X-ray position. The data were not sufficient to allow further conclusions regarding this source. Another source, 31669, was found to be closer to the position of the X-ray source than the counterpart suggested by Pooley et al. This source was only detected in our B-band images, so the data are not sufficient to draw any conclusions about its nature. However, based on the proximity to the X-ray source, we suggest that this may be the true counterpart.

\subsection{Other X-ray Sources}
\label{xray_other}

\subsubsection{CX\,4: A Known Dwarf Nova}

CX\,4 was shown to be variable by Bailyn et al. (1996), who determined a period of 5.1\,hours. It was identified as a U~Gem type DN by Kaluzny \& Thompson (2009). This source, number 7180 in our catalogue, was identified in every wave-band in this dataset. Analysis of the NUV data suggested a best-fitting period of 6.9\,hours, but the strength of the Lomb-Scargle peak was marginal. The complete NUV light curve is shown in Figure \ref{fig_lightcurves}.

\subsubsection{CX\,2, CX\,3, CX\,5, CX\,6, CX\,10, CX\,13, CX\,15: Cataclysmic Variables}

Of the remaining sources, CX\,3 and CX\,13 were outside our WFC3 field of view. For CV candidate sources CX\,2, CX\,5, CX\,6, CX\,10, and CX\,15, we were able to detect the sources that Pooley et al. suggest are the optical counterparts, but the data are not sufficient to draw any further conclusions regarding their characteristics. For CX\,6, we found a second source within the $Chandra$ error circle. This is source 34433, which was identified in our B-band image only and is included in Table \ref{xraytable} for completeness.

\subsubsection{CX\,11 and CX\,17: The Others}

We did not detect the counterpart to CX\,11 that Pooley et al. found. They suggest that this source is a CV or a background galaxy, neither of which are ruled out by our lack of detection.

Source CX\,17, which is thought to be an MSP or a background galaxy (Pooley et al. 2002) is outside the WFC3 field of view.

\subsubsection{CX\,9, CX\,14, CX\,18 and CX\,19: No Optical Counterpart}

Pooley et al. were unable to locate an optical counterpart for sources CX\,9, CX\,14, CX\,18 or CX\,19, and we were unable to identify a counterpart either.

Kaluzny \& Thompson (2009) claim a match to CX\,19, which is also visible in our observations, but this source is outside the $0\farcs2$ $Chandra$ error circle in our catalogue. Kaluzny \& Thompson do not state the size of the area they searched for counterparts. However, the source they found is $\approx0\farcs5$ from the $Chandra$ position and corresponds to our source 9889. While Kaluzny \& Thompson suggest a period of $0.11$\,days, we found no such period in our NUV light curve.

\section{A Search for Millisecond Pulsar Counterparts}
\label{msp}

There are five known millisecond pulsars (MSPs) in NGC\,6752 (D'Amico et al. 2002), of which 3 (PSR\,B, D and E) are inside the field of view of our WFC3 and STIS observations. All three of these are known to be isolated (D'Amico et al. 2002), and optical emission from the pulsars themselves has not yet been detected. We used the uncertainty in radio position from D'Amico et al. (2002) to compare the MSP positions to the positions of sources in our catalogue, and found no optical counterparts. We note that the position of X-ray source CX\,11 is consistent with that of PSR\,D, but we did not detect any optical sources within the search region for this source. Based on nearby sources that were detected in our study, we set an upper brightness limit on the MSP counterparts of $NUV_{STMAG}\approx22.5$\,mag.

\section{Variable Sources}
\label{variables}

In addition to searching for variability among optical counterparts to X-ray sources, we also carried out a general search for variability using sources detected in our FUV and NUV images. Using the method described in Section \ref{xray}, we identified sources where the reduced $\chi^{2}$ value for the FUV or NUV magnitudes was significantly higher than that of other sources of similar brightness. These sources are likely to be variable, and are highlighted as such in the catalogue (Table \ref{cattable}).

Again following the method explained in Section \ref{xray}, we then produced power spectra for the outliers and estimated the probability shuffling the positions of the data could produce a higher peak. Based on the number of outliers we investigated, we expect $\approx1$ source with an FAP value $<0.015$. It should be noted that the presence of power implies that a source has variability at a specific frequency, which does not necessarily correspond to periodic variability. For example, red noise power spectra exhibit variability on a range of frequencies, but the variability is not perdiodic. Aside from the DNe, we found three sources with FAP $<0.015$ and three more sources with FAP $\leqslant0.05$, but periodograms of these sources tend to have peaks at frequencies close to zero, suggesting that the power spectra in these cases are dominated by long-term trends.

\section{Dynamical Status}
\label{dynamical}

The dynamical status of a GC can be described in terms of its concentration parameter, $c=log_{10}(r_{t}/r_{c})$, where $r_{t}$ is the tidal radius and $r_{c}$ is the core radius. A concentration parameter of $c\gtrsim2-2.5$ is considered to be on the verge of collapsing, undergoing core-collapse, or post-core-collapse (Meylan \& Heggie 1997). There is, as yet, no clear way to distinguish between these phases, but fitting radial profiles to single-mass King models, which are characterised by values of $r_{t}$ and $r_{c}$, can give an indication of whether the entire cluster can be defined using one value of $c$, or if different values are needed to describe the core and outer regions.

Although NGC\,6752 is one of the closest GCs, its dynamical status is still a topic of debate. Djorgovski \& King (1986) and Auriere \& Ortolani (1989) suggested that it has undergone core-collapse and might be in a post-core-collapse bounce phase. Lugger et al. (1995), however, claimed that the radial profile is not inconsistent with a single King profile. Later, Ferraro et al. (2003) argued that two King profiles were needed to fit the radial distribution based on star counts, and that a post-core-collapse bounce is the most likely scenario. In a study of 38 GCs, Noyola \& Gebhardt (2006) found that NGC\,6752 was the only likely core-collapsed cluster to exhibit a flat core in the surface brightness profile. In this section we construct radial profiles based on our WFC3 and ACS data.

\subsection{Finding the Cluster Centre}
\label{clustercentre}

In order to construct radial profiles of the cluster, we first had to determine the location of the cluster centre. This is important,  because mis-placing the cluster centre will lead to a flattening of the radial density profile, hiding possible core structure. There are two distinct, basic ways to define the centre: using luminosity, $C_{lum}$, or using mass, $C_{grav}$. As stellar luminosity varies differently with stellar mass for different types of stars, these two points are not necessarily at the same location. Furthermore, different types of stars dominate the cluster's luminosity in different observational wave-bands, so $C_{lum}$ measured in one band may differ from that measured in another. It has been shown that $C_{grav}$ is the better measure of the `true' centre, in terms of consistency between colour bands (e.g. Montegriffo et al. 1995), and radial symmetry (e.g. Calzetti et al. 1993). In this study, therefore, we will use $C_{grav}$.\footnote{It is important to note, however, that the estimate of the centre that we refer to as centre of gravity, $C_{grav}$, is determined using the distribution of all stars, regardless of their mass or evolutionary status, so is not, strictly speaking, the gravitational centre. It is actually the geometrical centre.}

Following Dieball et al. (2010), we estimate the location of the cluster centre by finding the position at which the number of sources contained within a circular region of given radius $r_{lim}$ is a maximum. This was carried out using sources in our U-band catalogue, down to limiting magnitude of $U_{STMAG}=19$\,mag. Our U-band catalogue is particularly good for this, as it contains a sufficiently large number of sources (8546 sources have $U_{STMAG}<19$\,mag), but is not seriously affected by crowding. The adopted magnitude limit ensures that faint stars in the wings of bright sources are not missed, which would create a discrepancy between the centres found using only bright or only faint sources. For our final result, we used $r_{lim}=300$\,pixels $(=12\arcsec)$, but other reasonable choices gave results consistent with this. To estimate the uncertainties in our measurements, we used a simple bootstrapping method which involved sampling with replacement from the catalogue to create 1000 fake catalogues and estimating the centre of each one. The standard deviation of these `fake' centres gives the error on the real measurement.

As a result, we find the cluster centre to be at $x=2404\pm12$\,pixels, $y=2411\pm13$\,pixels in our local (WFC3) coordinate system. For comparison, we also used the V- and I-band source positions from the ACS Survey Catalogue. Using the same method we estimated the centre to be at $x=2404\pm13$, $y=2411\pm15$\,pixels in our local (WFC3) coordinate system, in excellent agreement with our U-band based estimate. This corresponds to $\alpha=19^{h}10^{m}52.128^{s}$, $\delta=-59^{\circ}59\arcmin04\farcs56$ in our Tycho-based system, with an estimated uncertainty of $0\farcs5$.

Table \ref{clustercentretable} lists estimates of the centre position from the literature. Ferraro et al. (2003) found the average $\alpha$ and $\delta$ coordinates of all stars in the PC chip of their WFPC2 image. Using WFPC2, PC images, Noyola \& Gebhardt (2006) estimated the location of the centre by dividing the area surrounding each assumed centre position into eight regions. They define the centre to be the point for which the standard deviation in the number of sources in the surrounding regions is lowest. This can be described as the point around which sources are most symmetrically distributed. Goldsbury et al. (2010) used the same ACS catalogue as us and fitted ellipses to contours of constant number density. The average centre of these ellipses was interpreted as the centre. Our estimate is consistent with these previous results.

\begin{table}
\caption{Estimates of the cluster centre position.}
\label{clustercentretable}
\begin{center}
\begin{tabular}{cccc}
\hline
\hline
R.A          & Decl.        & Uncertainty & Reference               \\
(hh:mm:ss)   & (deg:mm:ss)  & ($\arcsec$) &                         \\
\hline
19:10:52.128 & -59:59:04.56 & 0.5         & This paper              \\
19:10:52.040 & -59:59:04.64 & 0.5         & Ferraro et al. (2003)   \\
19:10:52.240 & -59:59:03.81 & 1.5         & Noyola \& Gebhardt (2006)\\
19:10:52.110 & -59:59:04.40 & 0.1         & Goldsbury et al. (2010) \\
\hline
\end{tabular}
\end{center}
\end{table}

\subsection{Stellar Density Profiles}
\label{profiles}

For the purposes of determining radial profiles for NGC\,6752, we rely on the V- and I-band data from the ACS Survey Catalogue, as this is the deepest available dataset. Repeating the process using our WFC3/NUV catalogue gave the same overall shape, with a simple shift in density that was consistent at all radii, indicating that depth is the dominant difference between the observed distributions.

The procedure we used is similar to that described by Djorgovski (1988). The catalogue is divided into concentric circles, centred on our cluster centre, increasing in radius by 50 pixels ($2\arcsec$) each step. Each annulus is split into eight equal sectors. Due to the off-centre location of the cluster centre on the chip, the number of segments which fell entirely on the chip in V- and I-band data ranges from 35 to 52, depending on the sector considered, while sectors in the NUV data contained 35 to 46 complete segments. The number of sources falling within a segment is divided by the area of the segment to give the stellar density.

The overall density for a given annulus is then computed as the average of the densities of the segments corresponding to that annulus. The error on the density of an annulus is the standard deviation on the mean density. This is justified by the fact that the dominant error source is the discrete nature of the distribution; if there is an unusually dense region of stars in one segment, that segment will contribute an artificially high number to the annulus, but the corresponding dispersion will also be larger.

In constructing the radial density profiles, we include all of the sources listed in the ACS Survey Catalogue (unlike in previous sections, in which only V- and I-band sources with a match in the WFC3 catalogues are included) This has a plate scale of $0\farcs05/$pixel out to radius $\simeq101\arcsec$. We limit the data to sources brighter than $\approx2$\,mags below the MSTO, corresponding to to $V_{STMAG}=19.7$\,mag, to reduce the likelihood of incompleteness impacting the results. We correct for completeness using the artificial star catalogue for NGC\,6752 that was created by Anderson et al. (2008) as part of the ACS Survey. Following their method, we consider a star to be recovered if the input and output fluxes agree to within 0.75\,mag and the positions agree to within 0.5\,pixels. Using the centre determined in Section \ref{clustercentre} we test for completeness as a function of radius, for stars brighter than $V_{STMAG}=19.7$\,mag. We find that completeness is almost 100\% at the edges of our images, and is over 50\% even in the core. We correct our stellar densities using the fractions of artificial sources recovered at each radius.

The radial profile obtained using this method is plotted in Figure \ref{fig_radialprofile}. The profile shows a continuing rise in density towards the centre of the cluster. This agrees with the findings of Ferraro et al. (2003), and is contrary to the findings of Noyola \& Gebhardt (2006), whose surface brightness profile showed a flat core within $log(r)\approx0\farcs5$. As shown in Table \ref{clustercentretable}, the centre used by Noyola \& Gebhardt differed from ours by nearly $1\arcsec$. Creating a radial density profile using our data but the centre given by Noyola \& Gebhardt, we find that the majority of the profile is unchanged, but the density at the innermost measurement point is slightly lower, giving the impression of a flatter core. The difference between our radial profile and that of Noyola \& Gebhardt could, therefore, be partly due to the difference in the assumed centre positions. An alternative explanation is that there is a fundamental difference between the stellar density profile and surface brightness profile. D'Amico et al. (2002) showed that NGC\,6752 has an usually high mass to light ratio ($\gtrsim10$), leading to a high proportion of optically faint sources in the core, which may contribute to the stellar density but not the surface brightness.

\subsection{Modelling the Stellar Density Profile}
\label{modelprofiles}

\begin{figure}
\includegraphics[width=0.5\textwidth]{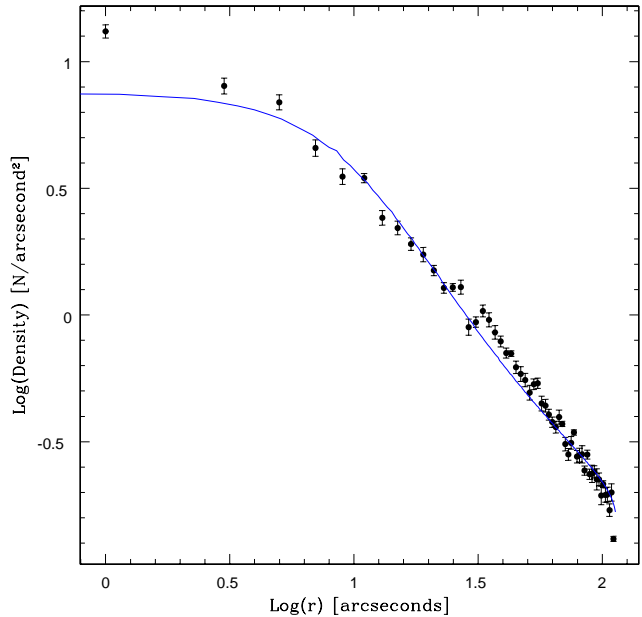}
\caption[fig_radialprofile]{Observed radial density profile using V- and I-band data from the ACS Survey Catalogue, based on the centre determined in Section \ref{clustercentre}. Blue, solid line: Best-fit King model to the data for the overall observed density profile, with $W_{0}=11$, $r_{c}=9\arcsec$, $c=2.547$. The poor fit to the data indicates that the profile of NGC\,6752 cannot be well modelled with a single King profile.}
\label{fig_radialprofile}
\end{figure}

In order to find the model which best describes the radial stellar density profile of NGC\,6752, we used single-mass King (1966) models, which we then projected to create 2-dimensional stellar density profiles. As we suspected that a plausible, single King model may not provide a good representation of the profile, we searched for the model that gave the best best fit to the data, without constraining the parameters to plausible physical ranges. Instead, we allowed the parameters ($W_{0}$, $r_{c}$, and $\sigma(0)$) to vary as required. The model that best described the data over the entire radius range had $W_{0}=11$, $r_{c}=9\arcsec$, $c=2.547$, and is plotted as a solid, blue line on Figure \ref{fig_radialprofile}. While this $r_{c}$ value is comparable to that of $10\farcs2$ given by Harris (1996, 2010 edition), a concentration parameter of $c\gtrsim2-2.5$ is usually considered to indicate a core-collapsed cluster, which should not be modelled using a single King profile. It is immediately apparent from the plot that the model does not give a good fit, demonstrating the difficulty in modelling the radial distribution of NGC\,6752 with a single King model, as found by Ferraro et al. (2003).

\begin{figure}
\includegraphics[width=0.5\textwidth]{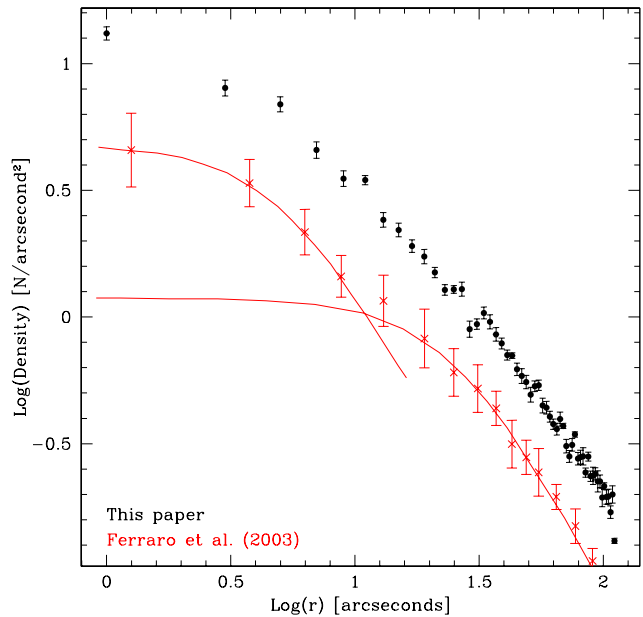}
\caption[fig_sara_ferraro]{Black points: Observed radial density profile using V- and I-band data from the ACS Survey Catalogue, based on the centre determined in Section \ref{clustercentre}. Red crosses/lines: Figure 5 of Ferraro et al. (2003), showing the two King models that they use to fit the data.}
\label{fig_sara_ferraro}
\end{figure}

As noted above, Ferraro et al. (2003) presented a radial profile of NGC\,6752 based on WFPC2 and ground based data, and also concluded that two King models were required to adequately fit the data. In their investigation, the PC chip of WFPC2 was roughly centred on the core of the GC, so the spatial resolution available was $0\farcs046/$pixel for the central region, out to $\approx18\farcs5$ radius. Outside this radius, the WFPC2/WF chips and ground based data were used, with pixel scale, at best, of $0\farcs1/$pixel. In the central region, the resolution of the images used by Ferraro et al. is comparable to that of ACS/WFC and WFC3, but beyond a radial distance of $\approx18\farcs5$, the dataset used in our investigation has considerably better resolution.

In Figure \ref{fig_sara_ferraro} we compare our data to the radial profiles from Ferraro et al. (2003), including the two King models they used to fit the data. The result shows that our measured V- and I-band stellar density is higher than theirs at all radii, because of the different brightness limits used (Ferraro et al. cut off at $V_{STMAG}=18.5$\,mag, while we include stars down to $V_{STMAG}=19.7$\,mag), but the shapes are quite similar.

\begin{figure}
\includegraphics[width=0.5\textwidth]{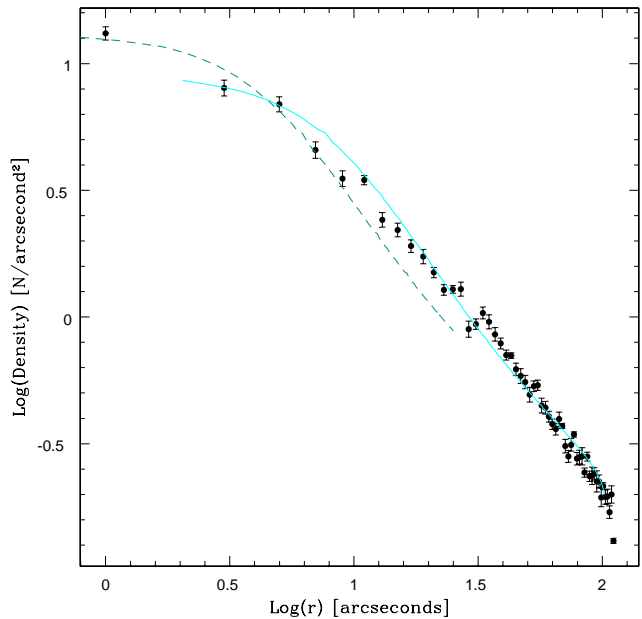}
\caption[fig_radialprofile_split]{Black points: Observed radial density profile using V- and I-band data from the ACS Survey Catalogue, based on the centre determined in Section \ref{clustercentre}. Dark green, dashed line: Best-fit King model to the innermost part of the cluster ($log_{10}(r)<0.9$). Cyan, solid line: Best-fit King model to the outer part of the cluster ($log_{10}(r)>1.2$). See text for details.}
\label{fig_radialprofile_split}
\end{figure}

As modelling the radial profile of NGC\,6752 with a single King model gave such a bad fit, we also split the data, following Ferraro et al. (2003), and used separate King profiles to fit the inner and outer parts of the cluster. We note that a double King model is a purely phenomenological structure, and has no actual physical basis. Furthermore, we treat the two parts of the cluster independently, neglecting any contribution to one regime from the other (for example, there should be a contribution to the inner part of the cluster's density profile from the underlying King model used in the outer part, which we have ignored). As before, we do not restrict the model parameters to physically reasonable ranges; instead we simply found the best fits to the data. We define the inner part of the cluster to have $log_{10}(r)<0.9$, and the outer part to have $log_{10}(r)>1.2$. Figure \ref{fig_radialprofile_split} shows the resulting fits. The inner part of the cluster (dark green, dashed line) is best described by a King model with $W_{0}=13$, $r_{c}=5\arcsec$, $c=2.944$, while the outer part (cyan, solid line) has $W_{0}=11$, $r_{c}=9\arcsec$, $c=2.547$. The core radius for the inner part is not dissimilar from that found by Ferraro et al. (2003). The core radius for the outer part is not a good match, but at $log_{10}(r)>1.2$ there is little difference in the shapes of models with different core radii. The values of $W_{0}$ and $c$ that we found are considerably higher than expected, but a large change in $W_{0}$ leads to a very small change in the shape of the profile, particularly near the core, and $c$ comes directly from the best-fit $W_{0}$ value. Models with lower values of $W_{0}$ (and, therefore, $c$) fit the data almost as well, especially at small radial distances.

Comparing Figure \ref{fig_radialprofile} and Figure \ref{fig_radialprofile_split}, it is clear that NGC\,6752's radial density profile is better fit using a combination of two King profiles than a single one, suggesting that the cluster is undergoing, or has undergone, a core-collapse phase.

\section{Radial Distributions and Masses of Stellar Populations}
\label{stellarpops}

\subsection{Radial Distributions}
\label{radialpops}

\begin{figure}
\includegraphics[width=0.5\textwidth]{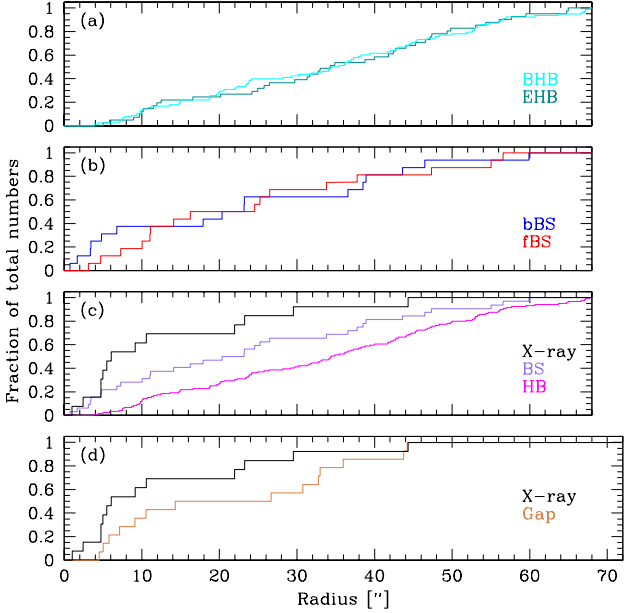}
\caption[fig_radial_pops]{Cumulative radial distributions for various stellar populations identified in our CMDs and of the X-ray sources within our field of view. Panels (a) and (b) compare the radial distributions of the EHB and BHB sources, and those of bright (blue) BSs and faint (red) BSs. Panel (c) shows the radial distributions of the X-ray sources, along with the (overall) BS and HB populations. Panel (d) compares the distribution of X-ray sources with that of the gap sources, down to a limiting magnitude of $NUV_{STMAG}=22.5$\,mag. See text for details.}
\label{fig_radial_pops}
\end{figure}

The cumulative radial profiles of various stellar populations identified in our CMDs and of the X-ray sources are shown in Figure \ref{fig_radial_pops}. In order to prevent completeness affecting the results, we use V- and I-band data to consider the HB stars and BSs, while the positions of gap sources are determined using measurements taken with NUV and U-band filters. We only consider sources brighter than $NUV_{STMAG}=22.5$\,mag to make sure that completeness does not affect the results. In this way, we can be confident that the sources have been categorised correctly and that the radial distributions are not biased by `missing' faint sources towards the core. We limit the distributions to a radial distance of $68\arcsec$ in the ACS (V- and I-band) data and $72\arcsec$ in WFC3 (NUV and U-band) data, in order to avoid bias due to the edges of the detectors. Only 13 of the 19 known X-ray sources are included, because the others are more than $72\arcsec$ from the core. Table \ref{radialpopstable} lists the number of sources considered from each stellar population.

Kolmogorov-Smirnov (KS) tests were carried out on various pairs of populations. The KS test calculates the probability that a difference in distribution as large as that observed can occur amongst sources drawn from the same underlying distribution. Therefore, the higher the percentage given by a KS test, the more likely it is that the two populations come from the same parent distribution, whereas a low percentage means that the two populations are significantly different. We caution that some of the samples used are relatively small, so care should be taken when interpreting the KS test results. Table \ref{kstable} shows the results of the KS tests.

Figure \ref{fig_radial_pops} shows that X-ray sources are the most centrally concentrated population, with BSs and gap sources also being centrally concentrated. This is to be expected, since all of these types of sources may be formed through dynamical interactions which are far more likely in the dense cluster core. Furthermore, X-ray binaries, gap sources (CVs and non-interacting MS-WD binaries) and BSs (MS-MS binaries; see Ferraro et al. 2004; Knigge et al. 2009) are expected to be more massive than ordinary cluster members, so will sink towards the core due to mass segregation. As shown in Table \ref{kstable}, there is no significant difference between the distributions of X-ray sources and gap sources or BSs (KS test results of 30.34\% and 10.48\%, respectively), but the X-ray sources have a significantly different radial distribution from the HBs (KS test shows that the likelihood of the X-ray and HB sources resulting from the same underlying distribution is 0.08\%). The distribution of gap sources (brighter than $NUV_{STMAG}=22.5$\,mag) is much closer to that of BSs than HBs, but the KS test does not give statistically significant results for either comparison. There is no evidence to suggest that the EHB and BHB sources (see panel (a) of Figure \ref{fig_radial_pops}) are formed from different underlying populations.

\begin{table}
\caption{Number of sources from each stellar population considered in computing radial profiles.}
\label{radialpopstable}
\begin{center}
\begin{tabular}{lc}
\hline
\hline
Population   & N$_{sources}$ \\
\hline
HB (overall) & 119 \\
BHB          & 78 \\
EHB          & 41 \\
BS (overall) & 32 \\
bBS          & 16 \\
fBS          & 16 \\
gap          & 14 \\
X-ray        & 13 \\
\hline
\end{tabular}
\end{center}
\end{table}

\begin{table}
\caption{Result of KS tests: probability in \% that a single population can exhibit differences in radial distributions as large as those observed. This gives an indication of the likelihood that the two populations are from the same distribution. Significant KS test results (i.e. KS probability $<5$\,\%) are highlighted in bold.}
\label{kstable}
\begin{center}
\begin{tabular}{lr}
\hline
\hline
        Populations   & KS test result \\
                      & [\%]           \\
\hline
        BHB vs. EHB   & 73.15          \\
        bBS vs. fBS   & 23.83          \\
\textbf{BS vs. HB}    & \textbf{ 4.13} \\
        BS vs. BHB    &  7.94          \\
        BS vs. EHB    &  5.21          \\
        BS vs. gap    & 91.69          \\
        BS vs. X-ray  & 10.48          \\
        gap vs. HB    &  9.21          \\
        gap vs. BHB   &  9.30          \\
        gap vs. EHB   & 18.49          \\
        gap vs. X-ray & 30.34          \\
\textbf{X-ray vs. HB} & \textbf{0.08}  \\
\textbf{X-ray vs. BHB}& \textbf{0.12}  \\
\textbf{X-ray vs. EHB}& \textbf{0.14}  \\
\hline
\end{tabular}
\end{center}
\end{table}

We also compare brighter and fainter BSs, following the method described in Dieball et al. (2010). Like them, we found that UV-bright BSs were also bluer in V-band (7 out of 7 sources with $NUV_{STMAG}<17$\,mag also have $M_{V-I}<-0.7$\,mag, compared to just 2 out of 25 with $NUV_{STMAG}>17$\,mag and $M_{V-I}<-0.7$\,mag). Bright BSs (bBSs) are thought to be younger (Ferraro et al. 2003b) and more massive (Sills et al. 2000) than faint BSs (fBSs), so should be more centrally concentrated (assuming that all BSs are older than the GC's relaxation time). Dieball et al. (2010) found, contrary to expectations, that fBSs were more centrally concentrated than bBSs. They suggested that BSs might get a kick at formation so that the younger, bBSs have not yet had time to sink back to the core. As shown in panel (b) of Figure \ref{fig_radial_pops}, the distribution in BSs in NGC\,6752 does not follow that of M\,80, but is more in line with the conventional model. There is no significant difference between the radial profiles of bright and faint BSs, and the KS test shows that there is no substantial evidence that they come from different initial distributions.

\subsection{Masses of Populations}
\label{masspops}

\begin{figure}
\includegraphics[width=0.5\textwidth]{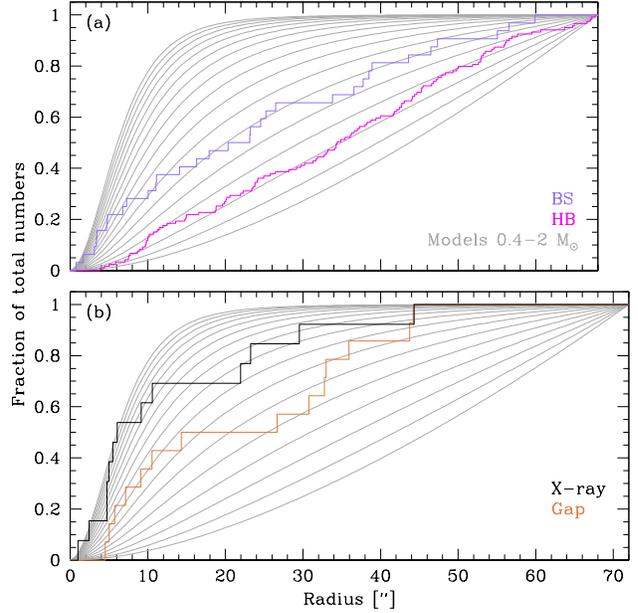}
\caption[fig_masses]{Comparison of the radial distributions of various stellar populations with theoretical King models with average masses of $0.4\,M_{\odot}$ (lowest grey line in each panel) to $2\,M_{\odot}$ (top grey line), in steps of $0.1\,M_{\odot}$. Panel (a) shows that the BS distribution agrees well with a mass of $0.9\,M_{\odot}$, while the HB population is consistent with a mass of $0.6\,M_{\odot}$. Panel (b) shows the radial distribution of X-ray sources and gap sources. The X-ray sources have masses larger than $1.1\,M_{\odot}$, while the gap sources have masses greater than $0.8\,M_{\odot}$.}
\label{fig_masses}
\end{figure}

The typical masses of stars belonging to a given stellar population can be estimated by comparing the populations' radial distribution to that of theoretical distributions of stars of a given mass. Using the method described in Heinke et al. (2003), we assume that the cluster can be well described by a classic King (1966) model (although as discussed in Section \ref{profiles}, this may be over-simplified), and compare the radial distributions of our sources to those of generalised theoretical King models described by
\[S(r)=\int{(1+(\frac{r}{r_{c\star}})^{2})^{\frac{1-3q}{2}}}dr,\]
where $r_{c\star}$ is the core radius and the parameter $q=M_{X}/M_{\star}$ is the ratio of the mass of the stellar population used to determine $r_{c\star}$ to the mass of the stellar population being considered. We take an MSTO star with mass $0.8\,M_{\odot}$ to be a typical star that defines the core radius and adopt the core radius determined by Trager, Djorgovski \& King (1993) using stars at or brighter than the MSTO of $r_{c}=10\farcs47$.

As in Section \ref{radialpops}, we use V- and I-band data to compare the HB stars and BSs, and NUV and U-band detections for the gap sources. Again, we only consider NUV sources brighter than $NUV_{STMAG}=22.5$\,mag. This ensures a uniform completeness limit. In order to avoid inconsistencies due to the edge of the field of view, we limit the area considered to a circle centred on the cluster core with radius $68\arcsec$ for ACS data and $72\arcsec$ for WFC3 data. We consider all of the X-ray sources that are within $72\arcsec$ of the cluster core.

Figure \ref{fig_masses} shows models for sources of mass $0.4-2M_{\odot}$, along with the radial distributions of the BSs, HBs, X-ray sources and gap sources. Panel (a) gives a mass estimate of $\approx0.9\,M_{\odot}$ for the BSs and $\approx0.6\,M_{\odot}$ for the overall HB population (BHB and EHB). These results are consistent with theoretical expectations. The ZAMS suggests BS masses of $0.95-1.65\,M_{\odot}$ and the ZAHB suggests an HB mass range of $0.49-0.66\,M_{\odot}$. In both cases, the majority of sources lie towards the lower mass (fainter) end of the sequence. Panel (b) shows that the X-ray sources have characteristic dynamical mass larger than $1.1\,M_{\odot}$, while the gap sources have dynamical masses of at least $0.8\,M_{\odot}$.

\section{Summary}
\label{conc}

We have analysed FUV images taken with STIS, and NUV, U- and B-band images taken with WFC3 on board \textit{HST} of the nearby, dense GC NGC\,6752. We matched our catalogues to the V- and I-band catalogue from the ACS Survey Catalogue, to produce a catalogue with a total of 39411 sources. The NUV-U CMD shows plentiful BS and HB populations, along with a number of `gap' sources in the region where we expect to find CVs and non-interacting WD - MS binaries. The images are also deep enough to reveal 360 WDs.

By comparing the positions of sources in our catalogue with those of known X-ray sources, we have found that two X-ray sources, CX\,1 and CX\,7 which were previously thought to be CV candidates are actually DNe. Prior to this study only one DN was known to exist in NGC\,6752. We have identified previously unknown optical counterparts to two X-ray sources. Another source which is just outside the X-ray position uncertainty of CX\,12, source 7581 in our catalogue, shows variability with a period of 4.1\,hours. One X-ray source, CX\,16, was thought to be a BY Dra source. We suggest an alternative optical source as the true counterpart. A search for variability revealed a number of potentially variable sources, which are indicated in our catalogue.

Finally, using the U-band and V-band catalogue, we estimate the position of the geometrical centre of the cluster, and use this centre to produce stellar density profiles of the cluster. Contrary to the surface density profile created by Noyola \& Gebhardt (2006), we do not find a flat core; this may be because of the different centre position used, or because the stellar density profile and surface brightness profile of NGC\,6752 are physically different. We conclude that the radial profile cannot be well modelled using a single King model, indicating that the cluster is undergoing, or has undergone, core collapse.

\section*{Acknowledgments}

We thank Tony Bird and Eva Noyola for helpful discussions.

Some/all of the data presented in this paper were obtained from the Multimission Archive at the Space Telescope Science Institute (MAST). STScI is operated by the Association of Universities for Research in Astronomy, Inc., under NASA contract NAS5-26555. Support for MAST for non-HST data is provided by the NASA Office of Space Science via grant NNX09AF08G and by other grants and contracts.

\label{lastpage}
\end{document}